\documentclass[aps,pra,floatfix,showpacs,preprint,superscriptaddress]{revtex4-2}
\usepackage{bm}
\usepackage{mathrsfs}
\usepackage{amsmath}
\usepackage{amssymb}
\usepackage{revsymb}
\usepackage{accents}
\usepackage{color}
\usepackage[force]{feynmp-auto}
\usepackage{hyperref}
\usepackage{cleveref}
\usepackage{wasysym}

\usepackage{graphicx}

\begin{document}
\title{Nonlinear Compton scattering and nonlinear Breit-Wheeler pair production including the damping of particle states}
\author{T. \surname{Podszus}}
\email{podszus@mpi-hd.mpg.de}
\affiliation{Max Planck Institute for Nuclear Physics, Saupfercheckweg 1, D-69117 Heidelberg, Germany}
\author{V. \surname{Dinu}}
\email{victordinu@yahoo.com}
\affiliation{Department of Physics, University of Bucharest, P.O. Box MG-11, Măgurele 077125, Romania}
\author{A. \surname{Di Piazza}}
\email{dipiazza@mpi-hd.mpg.de}
\affiliation{Max Planck Institute for Nuclear Physics, Saupfercheckweg 1, D-69117 Heidelberg, Germany}
\date{\today}

\begin{abstract}

In the presence of an electromagnetic background plane-wave field, electron, positron, and photon states are not stable, because electrons and positrons emit photons and photons decay into electron-positron pairs. This decay of the particle states leads to an exponential damping term in the probabilities of single nonlinear Compton scattering and nonlinear Breit-Wheeler pair production. In this paper we investigate analytically and numerically the probabilities of nonlinear Compton scattering and nonlinear Breit-Wheeler pair production including the particle states' decay. For this we first compute spin- and polarization-resolved expressions of the probabilities, provide some of their asymptotic behaviors and show that the results of the total probabilities are independent of the spin and polarization bases. Then, we present several plots of the total and differential probabilities for different pulse lengths and for different spin and polarization quantum numbers. We observe that it is crucial to take into account the damping of the states in order for the probabilities to stay always below unity and we show that the damping factors also scale with the intensity and pulse duration of the background field. In the case of nonlinear Compton scattering we show numerically that the total probability behaves like a Poissonian distribution in the regime where the photon recoil is negligible. In all considered cases, the kinematic conditions are such that the final particles momenta transverse to the propagation direction of the plane wave are always much smaller than the particles longitudinal momenta and the main spread of the momentum distribution on the transverse plane is along the direction of the plane-wave electric field.
\end{abstract}
 
\maketitle

\section{Introduction}

With the ongoing progress in laser technology toward higher intensities, lasers represent a promising tool to test QED in a regime where quantum effects induced by the laser field play an important role. In QED the vacuum is \emph{pictorially} described as being populated by fluctuations of virtual electron-positron pairs which can be polarized by a sufficiently large electromagnetic field. The so called ``critical'' field of QED $F_{\text{cr}}=m^2/|e|=1.3\times 10^{16}\ \text{V}/\text{cm}= 4.4 \times 10^{13}\ \text{G}$ (here $m$ and $e<0$ are the electron mass and charge, respectively, and we use units where $\epsilon_0=\hbar=c=1$) \cite{Landau_b_4_1982,Fradkin_b_1991,Dittrich_b_1985} determines the typical field scale of QED. In the presence of an electric field of the order of $F_{\text{cr}}$ the vacuum becomes unstable under electron-positron pair production and the interaction energy of a Bohr magneton with a magnetic field of strength $F_{\text{cr}}$ is of the order of $m$. The field strength $F_{\text{cr}}$ corresponds to a critical laser intensity $I_{\text{cr}} \sim 10^{29}\ \text{W}/\text{cm}^2$, which is far from being reached by available lasers. Indeed, today's record for the laser peak intensity $I_0$ is about $1.1 \times 10^{23}\ \text{W}/\text{cm}^2$ \cite{Yoon_2021}, and even upcoming laser facilities are aiming for intensities of the order of $I_0 \sim 10^{23}-10^{24}\ \text{W}/\text{cm}^2$ \cite{APOLLON_10P,ELI,CoReLS,Bromage_2019,XCELS}. However, due to the Lorentz-invariance of QED, physical observables like transition probabilities depend on the electromagnetic field only via Lorentz- and gauge-invariant parameters, such that the interesting regime where field-induced quantum effects dominate the dynamics can be efficiently entered experimentally already with today's technology. For an electron (photon) of four-momentum $p^{\mu}=(\epsilon,\bm{p})$ ($q^{\mu}=(\omega,\bm{q})$) moving in a background field, represented by the field tensor $F_0^{\mu\nu}=(\bm{E}_0,\bm{B}_0)$ in the laboratory frame, the probability of a physical process depends on the so-called quantum nonlinearity parameter $\chi_0=\sqrt{\left| (F_0^{\mu\nu}p_{\nu})^2\right| }/mF_{\text{cr}}$ ($\kappa_0=\sqrt{\left| (F_0^{\mu\nu}q_{\nu})^2\right| }/mF_{\text{cr}}$), with the metric tensor $\eta^{\mu\nu}=\text{diag}(+1,-1,-1,-1)$ \cite{Mitter_1975,Ritus_1985,Ehlotzky_2009,Reiss_2009,Di_Piazza_2012,Dunne_2014,Gonoskov_2021,Fedotov_2022}. In case of an electron or positron this parameter corresponds to the field strength that the electron or positron experiences in its rest frame in units of the critical field.

First experiments probing laser-electron interactions in the regime $\chi_0\lesssim 1$ were performed in the late 90s at SLAC \cite{Bula_1996,Burke_1997,Bamber_1999} and recently two experiments have been carried out close to the $\chi_0\sim 1$ regime by using an all-optical setup where the electron beam was generated via laser wake-field acceleration \cite{Cole_2018,Poder_2018}. Further experiments for testing the strong-field regime of QED with intense lasers are planned at DESY \cite{Abramowicz_2019} and at SLAC \cite{Meuren_2020}.

Already at intensities lower than $I_{\text{cr}}$ classical nonlinear effects due to the interaction with the background field become significant and complicate the theoretical description of the electron and positron dynamics. These nonlinear effects are controlled by the so-called classical nonlinearity parameter $\xi_0=|e|E_0/m\omega_0$, where $E_0$ is the electric field amplitude and $\omega_0$ is the central angular frequency of the laser pulse. For $\xi_0 \gtrsim 1$ the energy that an electron or a positron gains by the acceleration in the background field in one laser wavelength is comparable to its rest energy and the interaction of the charge with the background field cannot be treated perturbatively anymore \cite{Mitter_1975,Ritus_1985,Ehlotzky_2009,Reiss_2009,Di_Piazza_2012,Dunne_2014,Gonoskov_2021,Fedotov_2022}. Indeed, for optical lasers the parameter $\xi_0$ exceeds unity already at intensities of the order of $10^{18}\ \text{W}/\text{cm}^2$ and the interaction with the background field has to be treated exactly in the calculations in this case. This problem is commonly solved by working in the so-called Furry picture \cite{Furry_1951}. Here the background field is taken into account in the quantization procedure of the fermion field \cite{Fradkin_b_1991,Landau_b_4_1982}, such that the corresponding Dirac equation includes the interaction with the background field. An analytical solution of the Dirac equation can be found in the case of a plane-wave background field \cite{Volkov_1935} (see also Ref. \cite{Landau_b_4_1982}) and the corresponding states are known as Volkov states.

Two elementary processes which have been thoroughly investigated by employing Volkov states are the emission of a photon by an electron or positron (nonlinear Compton scattering) and the decay of a photon into an electron-positron pair (nonlinear Breit-Wheeler pair production), which, among others, were studied in Refs. \cite{Goldman_1964,Nikishov_1964,Ritus_1985,Baier_b_1998,Ivanov_2004,Boca_2009,
Harvey_2009,Mackenroth_2010,Boca_2011,Mackenroth_2011,Seipt_2011,Seipt_2011b,
Dinu_2012,Krajewska_2012,Dinu_2013,Seipt_2013,
Krajewska_2014,Wistisen_2014,Harvey_2015,Seipt_2016,Seipt_2016b,Angioi_2016,
Harvey_2016b,Angioi_2018,Di_Piazza_2018,Alexandrov_2019,Di_Piazza_2019,Ilderton_2019_b} (nonlinear Compton scattering) and Refs. \cite{Reiss_1962,Nikishov_1964,Narozhny_2000,Roshchupkin_2001,Reiss_2009,
Heinzl_2010b,Mueller_2011b,Titov_2012,Nousch_2012,Krajewska_2013b,Jansen_2013,
Augustin_2014,Meuren_2016,Di_Piazza_2019,King_2020} (nonlinear Breit-Wheeler pair production). 

Considering an electron (photon) moving in a plane wave laser pulse, it turns out that the leading-order in the fine-structure constant $\alpha=e^2/4\pi \approx 1/137$ total probability of nonlinear Compton scattering (nonlinear Breit-Wheeler pair production) exceeds unity for a sufficiently long total phase duration $\Phi_L$ of the laser pulse (or for a sufficiently large laser intensity). Probabilities larger than unity are of course unphysical and in contradiction with the unitarity of the $S$-matrix. Instead, for nonlinear Compton scattering it can be interpreted as the average number of photons emitted by the electron in the classical limit rather than as a probability \cite{Glauber_1951}. In general, from the QED perspective the reason behind this apparent contradiction is that higher-order loop corrections and processes become significant. This is intuitively clear for nonlinear Compton scattering since the emission of several photons by an electron due to nonlinear multiple Compton scattering processes becomes sizable with an increasing pulse length \cite{Loetstedt_2009,Seipt_2012,Mackenroth_2013,King_2015,Dinu_2019,Dinu_2020}.
A first investigation to taking into account these effects was carried out in Ref. \cite{Di_Piazza_2010} for $\chi_0 \lesssim 1$. The probability for an arbitrary number of consecutive incoherent photon emissions by an electron was calculated, taking into account the recoil at each photon emission. Each probability was then ``renormalized'' by imposing that the total probability of emitting either no photons or an arbitrary number of photons was unity. The renormalization ensures that the final probability of nonlinear (multiphoton) Compton scattering stays below unity even for large phase lengths of the laser pulse. These results where confirmed in Refs. \cite{Neitz_2013,Neitz_2014} by means of a kinetic approach, which also included the effects of nonlinear Breit-Wheeler pair production \cite{Neitz_2014}. In Refs. \cite{Neitz_2013,Neitz_2014} also inclusive quantities like average momenta are computed and an approach to obtain the momentum expectation values of an electron including multiple photon emissions and loops has been put forward in Ref. \cite{Torgrimsson_2021}.

In Ref. \cite{Tamburini_2021} the probability of an electron emitting an arbitrary number of photons was derived via the following considerations. The probability of an electron emitting $N$ photons was calculated by first combining the probability of the electron to emit $N-1$ photons until a certain time $t$, with the probabilities to emit one photon between $t$ and $t+dt$, with $dt$ being an infinitesimal positive time interval, and with the probability of emitting no photons from time $t+dt$ on, and finally by integrating the result over all possible times $t$ (a similar method was used in Ref. \cite{Aleksandrov_2021} to compute the pair-production yield as an observable to diagnose the intensity of the laser beam producing the pairs). Additionally, the photon recoil was taken into account. The obtained recursive equation for nonlinear multiphoton Compton scattering contains an exponential damping term describing the ``decay'' of the electron state by emitting a photon. This damping term depends on time and on the energy of the electron, and ensures that the total probability of emitting either no or an arbitrary number of photons is unity. The results in the regime $\chi_0\lesssim 1$ were in agreement with those in Ref. \cite{Di_Piazza_2010}.

\begin{figure}
\begin{center}
\includegraphics[width=1\columnwidth]{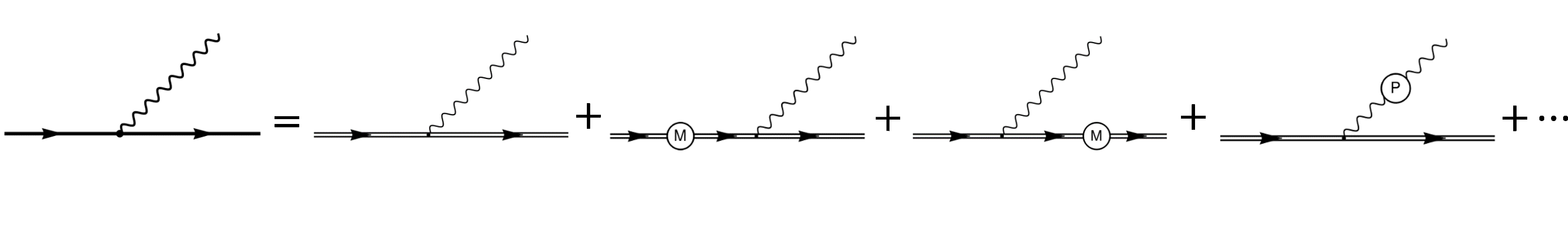}
\caption{The amplitude of nonlinear Compton scattering computed with the exact electron and photon states (thick straight and wiggly lines, respectively) is equal to the resummation of all one-particle reducible diagrams with corrections to the one-loop mass operator (M) on the Volkov-electron states (thin double lines) and to the one-loop polarization operator (P) on the photon states (thin wiggly lines) (see Ref. \cite{Podszus_2021}).}
\label{Figure_1}
\end{center}
\end{figure}

In Ref. \cite{Podszus_2021} the probability of nonlinear Compton scattering and of nonlinear Breit-Wheeler pair production was computed from first principles. The derived expressions are equivalent to the resummation of all one-particle reducible diagrams containing an arbitrary number of corrections to the electron and photon states by the one-loop mass and polarization operators, respectively (see Fig. \ref{Figure_1} for the case of nonlinear Compton scattering). This was achieved by calculating the $S$-matrix with the ``exact'' electron and photon states, which are the solutions of the Schwinger-Dyson equation. This work and Refs. \cite{Di_Piazza_2010,Neitz_2013,Neitz_2014,Torgrimsson_2021,Tamburini_2021} were framed within the so-called locally-constant-field approximation (LCFA). In the LCFA probabilities of QED processes reduce to the corresponding probabilities in a constant crossed field averaged over the phase-dependent plane-wave profile. This is reasonable in the limit of low-frequency plane waves with fixed electric-field amplitude, since here the formation length of QED processes is much smaller than the typical wavelength of the plane wave \cite{Ritus_1985}. The assumption is valid if $\xi_0 \gg 1$ at $\chi_0, \kappa_0 \sim 1$, which was assumed throughout the derivation (we did not consider the problems which occur for the LCFA at low photon energies in the case of nonlinear Compton scattering \cite{Di_Piazza_2018,Ilderton_2019_b}). It can be shown that the solution of the Schwinger-Dyson equation intrinsically includes the resummation of all corrections by the one-loop mass and polarization operator for the exact electron or positron and photon state, respectively. The final probabilities for nonlinear Compton scattering and nonlinear Breit-Wheeler pair production comprise an exponential damping term, describing the ``decay'' of the electron states by emitting a photon and the decay of the photon state into an electron-positron pair. This decay of the particle states turned out to become significant if $\alpha \xi_0 \Phi_L \gtrsim 1$. It is interesting to notice that the result for the probability of nonlinear Compton scattering is structurally similar to the single photon emission probability derived by the probabilistic approach in Ref. \cite{Tamburini_2021}. However, the probability in Ref. \cite{Podszus_2021} additionally includes spin and polarization effects as well as the decay of the photon state into an electron-positron pair.

In Ref. \cite{Dinu_2020} an approach has been developed to investigate higher-order QED processes also beyond the LCFA but for sufficiently long plane-wave pulses that the dynamics is dominated by the so-called cascade channel. In the cascade channel the higher-order process occurs as a sequence of the elementary building blocks represented by nonlinear Compton scattering and nonlinear Breit-Wheeler pair production.

The aim of the present work is to present new analytical insights and numerical examples on the probabilities of nonlinear Compton scattering and nonlinear Breit-Wheeler pair production including the particle states decay derived in Ref. \cite{Podszus_2021}. The paper is organized as follows. First, we give a detailed analytical evaluation of the differential and the total probabilities, together with the asymptotic behavior for one final particle getting all or none of the initial light-cone energy. Further, we prove that the results of the total probabilities are independent of the spin and polarization basis that we use in the calculations. Then, we pass to the numerical evaluation of the results. We present plots of the differential and the total probabilities at two different pulse lengths for both nonlinear Compton scattering and nonlinear Breit-Wheeler pair production. In the case of nonlinear Compton scattering we compare the results with a Poissonian distribution. For the differential probability we show different plots corresponding to different combinations of spin and polarization of the initial and final particles. Finally, the main conclusions of the paper are presented. An appendix contains the explicit computation of the traces of Dirac matrices, which are not essential for the understanding of the main results of the paper.

\section{Analytical calculations}

\subsection{Notation}

As indicated in the introduction, we investigate the probability of nonlinear Compton scattering and nonlinear Breit-Wheeler pair production including the particle states decay, which was presented in Ref. \cite{Podszus_2021} and for better comparison we employ the same notation here.

We consider a plane-wave background field with central photon four-momentum $k_0^{\mu}=\omega_0 n^{\mu}$, where we introduced the quantity $n^{\mu}=(1,\bm{n})$, with the three-dimensional unit vector $\bm{n}$ pointing along the direction of propagation of the background field. Additionally, we introduce the quantity $\tilde{n}^{\mu}=(1,-\bm{n})/2$ such that $(n \tilde{n})=1$. Since the background field is a plane wave, its four-potential $A^{\mu}(\phi)=(A^0(\phi),\bm{A}(\phi))$ only depends on the light-cone time $\phi=(n x)=t- \bm{n}\cdot \bm{x}$. Further, it is a solution of the free wave equation $\partial_{\mu} \partial^{\mu} A^{\nu}=0$, as we assume that it fulfills the Lorentz gauge condition $\partial_{\mu} A^{\mu} (\phi)=0$. By additionally fixing the gauge such that $A^0(\phi)=0$, the vector potential is perpendicular to the direction of propagation of the plane wave, i.e., $\bm{n} \cdot \bm{A}(\phi)=0$, if $\bm{A}(\phi)\to 0$ at $\phi\to \pm\infty$, which we also assume in the following.

By introducing the two four-vectors $a_j^{\mu}=(0,\bm{a}_j)$ with $j=1,2$, which obey the relations $(n a_j)=-2(\tilde{n}a_j)=-\bm{n} \cdot \bm{a}_j=0$ and $(a_j a_{j'})=-\bm{a}_j \cdot \bm{a}_{j'} =-\delta_{j j'}$, with $j,j'=1,2$, the vector potential of the plane wave can be expressed as $\bm{A}(\phi)=\psi_1(\phi) \bm{a}_1 +\psi_2(\phi) \bm{a}_2$. Here, $\psi_j(\phi)$ denotes the $j$th pulse shape function and it vanishes for $\phi\to\pm\infty$. In the following we will only consider the case of a linearly polarized plane wave, such that we choose without loss of generality $\psi_2(\phi)=0$ and $\psi_1(\phi)=A_0 \psi(\phi)$, with $A_0 <0$ being related to the amplitude of the electric field of the plane wave. The functions $\psi(\phi)$ and $\psi'(\phi)$ are assumed to be such that $|\psi(\phi)|\lesssim 1$ and $|\psi'(\phi)|\lesssim \omega_0$. Here and in the following a prime at a function denotes the derivative of the function with respect to its argument. The electromagnetic field tensor is then given by $F^{\mu\nu}(\phi)=n^{\mu} A'^{\nu}(\phi) -n^{\nu} A'^{\mu}(\phi)=A_0^{\mu\nu} \psi'(\phi)$, where we have introduced the notation $A_0^{\mu\nu}=A_0 (n^{\mu} a_1^{\nu}-n^{\nu} a_1^{\mu})$. The dual of the field tensor is $\tilde{F}^{\mu\nu}(\phi)=\tilde{A}_0^{\mu\nu} \psi'(\phi)$, where $\tilde{A}_0^{\mu\nu}=(1/2) \varepsilon^{\mu\nu\lambda\rho} A_{0,\lambda\rho}$ and we define the four-dimensional Levi-Civita tensor as $\varepsilon^{0123}=+1$ (note that in the chosen reference frame it is $\tilde{A}_0^{\mu\nu}=A_0 (n^{\mu} a_2^{\nu}-n^{\nu} a_2^{\mu})$). Since the four-potential and the field tensor always occur multiplied by the electron charge, we introduce the notation $\mathcal{A}^{\mu}(\phi)=e A^{\mu}(\phi)$, $\mathcal{A}_{0}=e A_{0}$, and $\mathcal{F}^{\mu\nu}(\phi)=e F^{\mu\nu}(\phi)$.

The quantities $n^{\mu}$, $\tilde{n}^{\mu}$, and $a_j^{\mu}$ fulfill the relation $\eta^{\mu \nu}=n^{\mu} \tilde{n}^{\nu} + \tilde{n}^{\mu} n^{\nu} - a_1^{\mu} a_1^{\nu} - a_2^{\mu} a_2^{\nu}$. It is useful to employ light-cone coordinates and for an arbitrary four-vector $v^{\mu}=(v_0,\bm{v})$ the components in light-cone coordinates are $v_-=(nv)=v_0-\bm{n}\cdot \bm{v}$, $v_+=(\tilde{n}v)=(v_0+\bm{n}\cdot \bm{v})/2$, and $\bm{v}_{\perp}=(v_{\perp,1},v_{\perp,2})=-((va_1),(va_2))=(\bm{v}\cdot\bm{a}_1,\bm{v}\cdot\bm{a}_2)$. Further we introduce the notation $\hat{v}=\gamma^{\mu} v_{\mu}$, where $\gamma^{\mu}$ are the Dirac-matrices, and we define $\gamma^5=i\gamma^0 \gamma^1 \gamma^2\gamma^3$.

\subsection{Nonlinear Compton Scattering}

In the case of nonlinear Compton scattering, we assume the incoming (outgoing) electron to have four-momentum $p^{\mu}=(\varepsilon,\bm{p})$ ($p'^{\mu}=(\varepsilon',\bm{p}')$), with energy $\varepsilon=\sqrt{m^2+\bm{p}^2}$ ($\varepsilon'=\sqrt{m^2+\bm{p}'^2}$), and an asymptotic spin quantum number $s=\pm 1$ ($s'=\pm 1$). The spin quantization axis of the incoming (outgoing) electron is chosen along the four-vector $\zeta^{\mu}=-\tilde{A}_0^{\mu\nu}p_{\nu}/(p_- A_0)$ ($\zeta^{\prime\,\mu}=-\tilde{A}_0^{\mu\nu}p'_{\nu}/(p'_- A_0)$), which corresponds to the three-dimensional spin vector $\bm{\zeta}$ ($\bm{\zeta}'$) pointing in the same direction of the magnetic field in the case of a constant crossed field and in the rest frame of the incoming (outgoing) electron. 
For the outgoing photon the four-momentum is $q^{\mu}=(\omega,\bm{q})$, with energy $\omega=|\bm{q}|$ and we define its two transverse polarization states, identified by the index $j=1,2$, along the four-vector $\Lambda_1^{\mu}(q)= A_0^{\mu\nu} q_{\nu}/(q_- A_0)$ and the pseudo-four-vector $\Lambda_2^{\mu}(q)= \tilde{A}_0^{\mu\nu} q_{\nu}/(q_- A_0)$, which fulfill the relation $(\Lambda_j(q) \Lambda_{j'}(q))=-\delta_{jj'}$ with $j,j'=1,2$.

With these definitions the probability of nonlinear Compton scattering including the damping of particle states within the LCFA, which was derived in Ref. \cite{Podszus_2021}, is given by the expression
\begin{equation}
\label{P_NCS}
\begin{split}
P^{(e^-\to e^-\gamma)}_{j,s,s'}=&\int\frac{d^3q}{16\pi^2}\frac{\alpha}{p_-p'_-\omega}\int d\phi_+e^{2\text{Im}\left\{\frac{m}{p_-}\int_{-\infty}^{\phi_+}d\varphi M_s(p,\varphi)+\int_{\phi_+}^{\infty}d\varphi \left[\frac{m}{p'_-}M_{s'}(p',\varphi)+\frac{m}{q_-}P_j(q,\varphi)\right]\right\}}\\
&\times\int d\phi_- e^{i\frac{m^2}{2p_-}\frac{q_-}{p'_-}\left\{[1+\bm{\pi}_{\perp,e}^2(\phi_+)]\phi_-+\frac{\bm{\mathcal{E}}^2(\phi_+)}{m^2}\frac{\phi_-^3}{12}\right\}} T_{j,s,s'},
\end{split}
\end{equation}
where we introduced the trace
\begin{equation}\label{Trace}
\begin{split}
T_{j,s,s'}=\frac{1}{4}\text{tr}&\left\{\left[1-\frac{\hat{n}[\hat{\mathcal{A}}(\phi_+)+\hat{\mathcal{A}}'(\phi_+)\phi_-/2]}{2p'_-}\right]\hat{\Lambda}_j(q)
\left[1+\frac{\hat{n}[\hat{\mathcal{\mathcal{A}}}(\phi_+)+\hat{\mathcal{A}}'(\phi_+)\phi_-/2]}{2p_-}\right]\right.\\
&\left.\times(\hat{p}+m)(1+s\gamma^5\hat{\zeta}) \left[1-\frac{\hat{n}[\hat{\mathcal{A}}(\phi_+)-\hat{\mathcal{A}}'(\phi_+)\phi_-/2]}{2p_-}\right]
\hat{\Lambda}_j(q)\right.\\
&\left.\times\left[1+\frac{\hat{n}[\hat{\mathcal{A}}(\phi_+)-\hat{\mathcal{A}}'(\phi_+)\phi_-/2]}{2p'_-}\right](\hat{p}'+m)(1+s'\gamma^5\hat{\zeta}')\right\},
\end{split}
\end{equation}
the transverse momentum
\begin{equation}\label{pi_e}
\bm{\pi}_{\perp,e}(\phi)=\frac{\bm{p}_{\perp}}{m}-\frac{p_-}{q_-}\frac{\bm{q}_{\perp}}{m}-\frac{\bm{\mathcal{A}}_{\perp}(\phi)}{m},
\end{equation}
and the plane-wave electric field (times the electron charge) $\bm{\mathcal{E}}(\phi)=-\bm{\mathcal{A}}'(\phi)$. Due to energy-momentum conservation the minus component and the perpendicular component of the outgoing electron momentum are fixed to $p'_-=p_- -q_-$ and $\bm{p}'_{\perp}=\bm{p}_{\perp} -\bm{q}_{\perp}$, respectively.
As we explained in the introduction, the probability comprises a damping term due to the particle states decay, which is the first exponential function in Eq. (\ref{P_NCS}). The exponent contains the imaginary parts of the mass and polarization operator, more precisely, the one-loop mass operator within the LCFA, given by the expression \cite{Baier_1976_a,Ritus_1970,Meuren_2011}
\begin{equation}
\label{M_s}
M_s(p,\phi)=\frac{\alpha m}{2\pi}\int_0^{\infty} du\int_0^{\infty}\frac{dv}{(1+v)^3}e^{-iu\left[1+\frac{1}{3}\frac{\chi^2_p(\phi)}{v^2}u^2\right]}\left[\frac{5+7v+5v^2}{3}\frac{\chi^2_p(\phi)}{v^2}u-is\chi_p(\phi)\right],
\end{equation}
for $s=\pm 1$, with the $\phi$-dependent quantum nonlinearity parameter defined as $\chi_p(\phi)=-(p_-/m) A_0 \psi'(\phi)/F_{\text{cr}}$, and the transverse part of the one-loop polarization operator within the LCFA, given by \cite{Meuren_2015,Baier_1976_b,Becker_1975}
\begin{equation}
\label{P_j}
P_j(q,\phi)=\frac{\alpha}{48\pi}m j\kappa_q^2(\phi)\int_0^{\infty}du\,u\int_0^1 dv e^{-iu\left[1+\frac{(1-v^2)^2}{48}\kappa_q^2(\phi)u^2\right]}(1-v^2)[3-(-1)^jv^2],
\end{equation}
for $j=1,2$, with $\kappa_q(\phi)=-(q_-/m) A_0 \psi'(\phi)/F_{\text{cr}}$. 

According to the optical theorem, the quantity $-(2m/p_-) \text{Im} [M_s (p,\phi)]$ is equal to the total probability per unit $\phi$ that an electron of four-momentum $p$ and spin quantum number $s$ emits a photon \cite{Baier_1976_a} and the quantity $-(2m/q_-) \text{Im} [P_j (q,\phi)]$ is equal to the total probability per unit $\phi$ that a photon of four-momentum $q$ and polarization quantum number $j$ decays into an electron-positron pair \cite{Baier_1976_b}.
Hence, this exponential damping can be understood as the electron and photon states not being stable in the background field but decaying, where the electron ``decays'' by emitting a photon and the photon decays into an electron-positron pair. 

Further, we notice that the damping exponential depends on the spin of the incoming and outgoing electrons and on the polarization of the outgoing photon. This prevents one from employing the commonly used spin and polarization sum rules when solving the trace in Eq. (\ref{Trace}), such that the spin- and polarization-resolved traces have to be calculated. A complete analytical derivation can be found in the appendix. Here, we only present the main steps of the calculations. With our choice of the spin and polarization basis, the trace for the two polarization states $j=1,2$ reduces in a linearly-polarized field to
\begin{equation}\begin{split}\label{T_1}
T_{1,s,s'} &= (1+s s') \bigg[ (p p')-m^2 
- \frac{1}{2} \frac{q_-}{p_-} \frac{q_-}{p_--q_-} \left(p_1-\frac{p_-}{q_-} q_1 \right)^2\\
& -\left( 2 +\frac{1}{2} \frac{q_-}{p_-} \frac{q_-}{p_--q_-}\right) \mathcal{A}_0^2 \psi'^2(\phi_+) \frac{\phi_-^2}{4} \\
&+\left( 2+\frac{1}{2} \frac{q_-}{p_-} \frac{q_-}{p_--q_-}\right) \left(p_1-\frac{p_-}{q_-} q_1 +\mathcal{A}_0 \psi(\phi_+) \right)^2 \bigg]\\
&+i (s+s') \frac{m}{2} \mathcal{A}_{0} \psi'(\phi_+) \phi_- \frac{q_-}{p_-}\left(2+\frac{q_-}{p_--q_-} \right)
- s s' \frac{q_-}{p_-} \frac{q_-}{p_--q_-} \left(p_2-\frac{p_-}{q_-} q_2 \right)^2
\end{split}\end{equation}
and 
\begin{equation}\begin{split}\label{T_2}
T_{2,s,s'} &= (1-s s') \bigg[ (p p')-m^2 
- \frac{1}{2} \frac{q_-}{p_-} \frac{q_-}{p_--q_-} \left(p_1-\frac{p_-}{q_-} q_1 \right)^2 \\
&- \frac{1}{2} \frac{q_-}{p_-} \frac{q_-}{p_--q_-} \mathcal{A}_0^2 \psi'^2(\phi_+) \frac{\phi_-^2}{4} \\
&+\frac{1}{2} \frac{q_-}{p_-} \frac{q_-}{p_--q_-} \left(p_1-\frac{p_-}{q_-} q_1 +\mathcal{A}_0 \psi(\phi_+) \right)^2 \bigg]\\
&+(1+ss') 2 \left(p_2-\frac{p_-}{q_-} q_2 \right)^2 + s s' \frac{q_-}{p_-} \frac{q_-}{p_--q_-} \left(p_2-\frac{p_-}{q_-} q_2 \right)^2\\
&-i (s-s') \frac{m}{2} \mathcal{A}_{0} \psi'(\phi_+) \phi_- \frac{q_-}{p_-}\frac{q_-}{p_--q_-}.
\end{split}\end{equation}
We observe that the traces depend on the pulse shape function $\psi(\phi_+)$. However, it should be possible to express the dependence on $\psi(\phi_+)$ in a manifestly gauge-invariant way. In order to achieve this, we consider the transverse momentum $\bm{\pi}_{\perp,e}(\phi_+)$ defined in Eq. (\ref{pi_e}). After some simplifications all the dependence on $\psi(\phi_+)$ and $q_1$ turns out to be in the electron quasi-momentum $\bm{\pi}_{\perp,e}(\phi_+)$, which can be removed by performing the integral in $\phi_-$ and using the properties of the Airy functions. Indeed, terms proportional to the derivative in $\phi_-$ of the second exponential function in Eq. (\ref{P_NCS}), i.e., terms proportional to $\left[1+\bm{\pi}^2_{\perp,e}(\phi_+)+(\mathcal{A}_0^2\psi'^2(\phi_+)/m^2)(\phi_-^2/4)\right]$, vanish when performing the integral over $\phi_-$. By performing the appropriate substitutions, the traces ultimately depend only on the derivative $\psi'(\phi_+)$ of the pulse shape function and the probability is therefore manifestly gauge-invariant. Ignoring the corresponding vanishing terms, the traces can be equivalently written as
\begin{equation}\begin{split}\label{T_prime_1}
T_{1,s,s'} &= -2 (1+s s') m^2 -(1+s s') \left( 4+ \frac{q_-}{p_-} \frac{q_-}{p_--q_-}\right) \mathcal{A}_0^2 \psi'^2(\phi_+) \frac{\phi_-^2}{4}\\
&+i (s+s') \frac{m}{2} \mathcal{A}_{0} \psi'(\phi_+) \phi_- \frac{q_-}{p_-}\left(2+\frac{q_-}{p_--q_-} \right)\\
&-\left(2+2ss'+ s s' \frac{q_-}{p_-} \frac{q_-}{p_--q_-}\right) \left(p_2-\frac{p_-}{q_-} q_2 \right)^2
\end{split}\end{equation}
and 
\begin{equation}\begin{split}\label{T_prime_2}
T_{2,s,s'} &= -(1-s s') \frac{q_-}{p_-} \frac{q_-}{p_--q_-} \mathcal{A}_0^2 \psi'^2(\phi_+) \frac{\phi_-^2}{4}\\
&-i (s-s') \frac{m}{2} \mathcal{A}_{0} \psi'(\phi_+) \phi_- \frac{q_-}{p_-}\frac{q_-}{p_--q_-}\\
&+\left(2+2ss'+ s s' \frac{q_-}{p_-} \frac{q_-}{p_--q_-}\right) \left(p_2-\frac{p_-}{q_-} q_2 \right)^2.
\end{split}\end{equation}
We transform the integral in the photon momentum into light-cone coordinates using the relation $d^3 q =(\omega/q_-) dq_- d^2q_{\perp}$ and introduce the notation
\begin{equation}\label{T_notation}
\tilde{T}_{j,s,s'}=-\frac{1}{4\pi^2 m^2} \frac{p_-}{q_- p'_-} \int d\phi_- \int d^2q_{\perp} e^{i\frac{m^2 q_-}{2p_- p'_-}\left\{[1+\bm{\pi}_{\perp,e}^2(\phi_+)]\phi_-+\frac{\bm{\mathcal{E}}^2(\phi_+)}{m^2}\frac{\phi_-^3}{12}\right\}} T_{j,s,s'}.
\end{equation}
Now, the integral in the perpendicular photon momentum $\bm{q}_{\perp}$ can be computed analytically by using the two basic integrals \cite{NIST_b_2010}
\begin{align}
&\int d^2 q_{\perp} e^{i \frac{m^2 q_-}{2p_- p'_-} \bm{\pi}^2_{\perp,e} (\phi_+) \phi_-} =2\pi i \frac{q_- p'_-}{p_- (\phi_- +i0)},\\
&\int d^2 q_{\perp} \left(p_2-\frac{p_-}{q_-} q_2 \right)^2 e^{i \frac{m^2 q_-}{2p_- p'_-} \bm{\pi}^2_{\perp,e} (\phi_+) \phi_-} = -2\pi \frac{ p'^2_-}{ (\phi_- +i0)^2} .
\end{align}

For the integral in $\phi_-$ we use the integral representation $\text{Ai}(z)=\int_{-\infty}^{\infty} (d\tilde{\phi}/2\pi) \exp (iz\tilde{\phi}+i\tilde{\phi}^3/3)$ of the Airy function \cite{NIST_b_2010}. With the substitutions $\tilde{\phi}=\left[q_- \bm{\mathcal{E}}^2(\phi_+)/(8p_- p'_-) \right]^{1/3} \phi_-$ and $z=\left[q_-/(p'_- \chi_p(\phi_+)) \right]^{2/3}$, the integral in $\phi_-$ can be taken. Hence, the probability of nonlinear Compton scattering including the damping of particle states is finally given by
\begin{equation}
\label{P_NCS_f}
\begin{split}
P^{(e^-\to e^- \gamma)}_{j,s,s'} = &
- \frac{\alpha m^2}{4 p_-^2} \int_0^{p_-} dq_- \int d\phi_+\ \tilde{T}_{j,s,s'} \\
&\times e^{2\text{Im} \left\{\frac{m}{p_-} \int_{-\infty}^{\phi_+} d\varphi M_s(p,\varphi) +\int_{\phi_+}^{\infty} d\varphi \left[ \frac{m}{p'_-} M_{s'}(p',\varphi) +\frac{m}{q_-} P_j(q,\varphi) \right] \right\} } ,
\end{split}
\end{equation}
where
\begin{equation}
\label{T_NCS_f1}
\begin{split}
\tilde{T}_{1,s,s'}
= &  \left[ 1+ss' \left(1-\frac{q_-^2}{2p_-(p_--q_-)}\right)\right] \text{Ai}_1(z) \\
&+\left[ 3+\frac{q_-^2}{p_-(p_--q_-)} +ss' \left(3+\frac{q_-^2}{2p_-(p_--q_-)}\right)\right] \frac{\text{Ai}'(z)}{z}\\
&+(s+s') \left(2\frac{q_-}{p_-}+ \frac{q_-^2}{p_-(p_--q_-)} \right) \frac{\text{Ai}(z)}{\sqrt{z}} \text{sgn}(\psi'(\phi_+)) 
\end{split}\end{equation}
and 
\begin{equation}
\label{T_NCS_f2}
\begin{split}
\tilde{T}_{2,s,s'}
= &  \left[ 1+ss' \left(1+\frac{q_-^2}{2p_-(p_--q_-)}\right)\right] \text{Ai}_1(z) \\
&+\left[ 1+\frac{q_-^2}{p_-(p_--q_-)} +ss' \left(1-\frac{q_-^2}{2p_-(p_--q_-)}\right)\right] \frac{\text{Ai}'(z)}{z}\\
&+(s'-s) \frac{q_-^2}{p_-(p_--q_-)} \frac{\text{Ai}(z)}{\sqrt{z}} \text{sgn}(\psi'(\phi_+)),
\end{split}\end{equation}
with $\text{Ai}_1(z)=\int_z^{\infty} dx \text{Ai}(x)$ and with $\text{sgn}(\psi'(\phi_+))$ denoting the sign of $\psi'(\phi_+)$. Note that without the exponential damping term the results reduce to the expressions of the spin- and polarization-resolved probabilities of nonlinear Compton scattering, which can be found in Ref. \cite{Seipt_2020}. This observation can be used to prove analytically that the probability $P^{(e^-\to e^- \gamma)}_{s}=\sum_{j,s'}P^{(e^-\to e^- \gamma)}_{j,s,s'}$ is always smaller than unity. In fact, since the damping exponentials are smaller or equal to unity, it is
\begin{equation}
\begin{split}
P^{(e^-\to e^- \gamma)}_s &< 
- \frac{\alpha m^2}{4 p_-^2} \sum_{j,s'}\int_0^{p_-} dq_- \int d\phi_+\ \tilde{T}_{j,s,s'} e^{2\text{Im}\frac{m}{p_-} \int_{-\infty}^{\phi_+} d\varphi M_s(p,\varphi)  }\\
&=\int d\phi_+ \frac{\partial P^{\text{NC}}_{s,p}}{\partial \phi_+} e^{-\int_{-\infty}^{\phi_+} d\varphi \frac{\partial P^{\text{NC}}_{s,p}}{\partial \varphi}   }=-\int d\phi_+ \frac{\partial}{\partial \phi_+} e^{-\int_{-\infty}^{\phi_+} d\varphi \frac{\partial P^{\text{NC}}_{s,p}}{\partial \varphi}   }\\
&=1-e^{-\int_{-\infty}^{\infty} d\varphi \frac{\partial P^{\text{NC}}_{s,p}}{\partial \varphi} }<1,
\end{split}
\end{equation}
where $\partial P^{\text{NC}}_{s,p}/\partial \phi$ indicated the probability without damping of nonlinear Compton scattering per unit of light-cone time $\phi$.

At this point we also want to investigate the asymptotic behavior of the differential probability for the two cases $q_-\ll p_-$ and $p_- -q_-\ll p_-$. In the first case, the photon recoil is negligible, whereas in the second case, almost all light-cone energy of the incoming electron goes into the photon. The differential probability is obtained from Eq. (\ref{P_NCS_f}) and it is given by 
\begin{equation}\label{diff_prob}
\begin{split}
\frac{\partial P^{(e^-\to e^-\gamma)}_{j,s,s'}}{\partial q_-} = &
- \frac{\alpha m^2}{4 p_-^2} \int d\phi_+ e^{D^{\text{NC}}_{j,s,s'} } \tilde{T}_{j,s,s'} ,
\end{split}\end{equation}
where we have renamed the exponent of the exponential damping function as
\begin{equation}\label{c_damp}
\begin{split}
D^{\text{NC}}_{j,s,s'}= 2\text{Im} \left\{\frac{m}{p_-} \int_{-\infty}^{\phi_+} d\varphi M_s(p,\varphi) +\int_{\phi_+}^{\infty} d\varphi \left[ \frac{m}{p'_-} M_{s'}(p',\varphi) +\frac{m}{q_-} P_j(q,\varphi) \right] \right\} .
\end{split}\end{equation}
As already mentioned, according to the optical theorem, the probabilities of nonlinear Compton scattering and nonlinear Breit-Wheeler pair production are related to the imaginary part of the mass and polarization operator, respectively. Hence, the exponent of the damping term, $D^{\text{NC}}_{j,s,s'}$, is equal to minus the sum of the probability of nonlinear Compton scattering between $-\infty$ and $\phi_+$ for the incoming electron with light-cone energy $p_-$ and spin quantum number $s$, and of the probabilities of nonlinear Compton scattering of the outgoing electron with light-cone energy $p'_-$ and spin quantum number $s'$ and of nonlinear Breit-Wheeler pair production of the outgoing photon with light-cone energy $q_-$ and polarization quantum number $j$, both between $\phi_+$ and $+\infty$ \cite{Baier_1976_a,Baier_1976_b}, i.e.,
\begin{equation}
\begin{split}\label{c_damp_prob}
D^{\text{NC}}_{j,s,s'}= -\int_{-\infty}^{\phi_+} d\varphi \frac{\partial P^{\text{NC}}_{s,p}}{\partial \varphi} -\int_{\phi_+}^{\infty} d\varphi \left( \frac{\partial P^{\text{NC}}_{s',p'}}{\partial \varphi} +\frac{\partial P^{\text{NBW}}_{j,q}}{\partial \varphi} \right).
\end{split}\end{equation}

\subsubsection{\textbf{Asymptotic expression for $q_- \ll p_-$}}

We first analyze the asymptotic expression of the differential probability in the asymptotic region $q_-\ll p_-$. Also, we assume that the quantum nonlinearity parameter $\chi_p(\varphi)$ of the electron is fixed, such that the absolute value of the quantum nonlinearity parameter $\kappa_q (\varphi)=(q_-/p_-)\chi_p(\varphi)$ of the photon is much smaller than unity (if $|\chi_p(\varphi)|$ is larger than unity, the ratio $q_-/p_-$ is assumed to be sufficiently small that $|\kappa_q (\varphi)|\ll 1$). Thus, we can use in the damping term the corresponding asymptotic expression for the probability of nonlinear Breit-Wheeler pair production: \cite{Ritus_1985}  
\begin{equation}
\frac{\partial P^{\text{NBW}}_{j,q}}{\partial \varphi} \overset{\kappa_q(\varphi) \ll 1}{\approx} \sqrt{\frac{3}{2}} \frac{\alpha m^2 |\kappa_q(\varphi)| j}{8q_-} e^{-\frac{8}{3|\kappa_q(\varphi)|}}. 
\end{equation}
Since it is exponentially suppressed in the limit $\kappa_q (\varphi)\to 0$, we neglect it below. Furthermore, due to the conservation of the minus component of the four-momentum, if $q_-\ll p_-$ then $p'_-\approx p_-$, and the damping function reduces to
\begin{equation}
\begin{split}\label{damp_NCS_class}
D^{\text{NC}}_{j,s,s'} \overset{q_-\ll p_-}{\approx} \frac{2m}{p_-} \text{Im} \left[ \int_{-\infty}^{\phi_+} d\varphi M_s(p,\varphi) +\int_{\phi_+}^{\infty} d\varphi M_{s'}(p,\varphi) \right] .
\end{split}\end{equation}
Now, we will shortly see that for $q_-\ll p_-$ the spin-dependent terms of the probability of nonlinear Compton scattering can be neglected \cite{Baier_b_1998}, such that the damping exponent effectively reduces to a constant, which is equal to minus the probability of an electron of momentum $p_-$ emitting a photon between phase $-\infty$ and $+\infty$ averaged over the electron spin.

In the functions $\tilde{T}_{j,s,s'}$ we can expand the Airy functions for $z=\left( \frac{q_-}{p'_- \chi_p(\phi_+)}\right)^{2/3}\approx\left( \frac{q_-}{p_- \chi_p(\phi_+)}\right)^{2/3} \ll 1$ (we assume that $\chi_p(\phi_+)$ is fixed and that the ratio $q_-/p_-$ is much smaller than $1/|\chi_p(\varphi)|$ if $|\chi_p(\varphi)|<1$) and we obtain
\begin{align}
\tilde{T}_{1,s,s} &\overset{z\ll 1}{\approx}  -\frac{2\times 3^{2/3}}{ \Gamma\left(\frac{1}{3}\right) z}, && \tilde{T}_{1,s,-s} \overset{z\ll 1}{\approx} - \frac{q_-^2}{2 p^2_-}\frac{1}{3^{1/3} \Gamma\left(\frac{1}{3}\right) z},\\
\tilde{T}_{2,s,s} &\overset{z\ll 1}{\approx}   -\frac{2}{3^{1/3} \Gamma\left(\frac{1}{3}\right) z}, && \tilde{T}_{2,s,-s} \overset{z\ll 1}{\approx} -\frac{q_-^2}{2 p^2_- }\frac{3^{2/3}}{ \Gamma\left(\frac{1}{3}\right) z}.
\end{align}
As expected in the classical limit where the photon recoil is small, we see that the probability of spin flip is substantially suppressed as compared to the case $s=s'$. Further, the damping function in Eq. (\ref{damp_NCS_class}) in the dominant case $s=s'$ is independent of the integration variables and corresponds in the classical limit to the mean number of photons emitted by an electron. Hence, within this limit the total probability of emitting a single photon is in agreement with that obtained from the Poissonian distribution.

\subsubsection{\textbf{Asymptotic expression for $p_--q_- \ll p_-$}}

Now, we evaluate the asymptotic expression of the differential probability in Eqs. (\ref{diff_prob}) and (\ref{c_damp}) in the region $p'_-=p_--q_-\ll p_-$. In this case, since we again assume $\chi_p (\varphi)$ to be fixed, the absolute value of the quantum nonlinearity parameter $\chi_{p'} (\varphi)=(p'_-/p_-)\chi_p (\varphi)$ of the outgoing electron is assumed to be smaller than unity. We use the corresponding asymptotic expression for the probability of nonlinear Compton scattering in Eq. (\ref{c_damp_prob}), which is independent of $p'$ and given by \cite{Ritus_1985}
\begin{equation}\label{NCS_chi_small}
\frac{\partial P^{\text{NC}}_{s',p'}}{\partial \varphi}\overset{\chi_{p'}(\varphi) \ll 1}{\approx} \frac{5}{2\sqrt{3}} \frac{\alpha m^2 |\chi_p(\varphi)|}{p_-}. 
\end{equation}
The damping function is then
\begin{equation}
\begin{split}
D^{\text{NC}}_{j,s,s'} \overset{p'_- \ll p_-}{\approx} 2\,\text{Im} \left[\frac{m}{p_-} \int_{-\infty}^{\phi_+} d\varphi M_s(p,\varphi) +\frac{m}{p_-} \int_{\phi_+}^{\infty} d\varphi P_j(p,\varphi) \right] - \int_{\phi_+}^{\infty} d\varphi \frac{5}{\sqrt{3}} \frac{\alpha m^2 |\chi_p(\varphi)|}{p_-} .
\end{split}
\end{equation}
In the functions $\tilde{T}_{j,s,s'}$ we assume that the ratio $p'_-/p_-$ is sufficiently small that $z=\left( \frac{q_-}{p'_- \chi_p(\phi_+)}\right)^{2/3}\approx\left( \frac{p_-}{p'_- \chi_p(\phi_+)}\right)^{2/3} \gg 1$. We obtain for photon polarization $j=1$ and identical spin quantum numbers $(s=s')$
\begin{equation}\begin{split}
\tilde{T}_{1,s,s} \overset{z\gg 1}{\approx}& -\frac{1}{\sqrt{\pi}} z^{-3/4} e^{-\frac{2}{3} z^{3/2}} \bigg[ \frac{p_-}{p'_- } \left(1 - s\ \text{sgn}(\psi'(\phi_+)) \right) + 2  \frac{p'_-}{p_-} s\ \text{sgn}(\psi'(\phi_+))   \bigg]\\
& - \frac{1}{96\sqrt{\pi}} z^{-9/4} e^{-\frac{2}{3} z^{3/2}} \left( 124 + 20 s\ \text{sgn}(\psi'(\phi_+)) \right)\\
& -\frac{1}{9216 \sqrt{\pi}} z^{-15/4} e^{-\frac{2}{3} z^{3/2}}  \frac{p_-}{p'_- } \left( 3938 - 770 s\ \text{sgn}(\psi'(\phi_+)) \right).
\end{split}\end{equation}
Note that here in the special case of $s=\text{sgn}(\psi'(\phi_+))$ compensations occur and this is why higher-order terms have been reported in the expression above. However, by keeping in mind that the functions $\tilde{T}_{j,s,s'}$ are ultimately integrated over the light-cone time $\phi_+$ to compute the emission probability, the function $\text{sgn}(\psi'(\phi_+))$ takes both values $+1$ and $-1$ (we implicitly assume here that the plane wave describes an oscillating laser wave). Therefore, the scaling of the probability will be determined by the term in $\tilde{T}_{1,s,s}$ scaling as $z^{-3/4}/p'_-$ and we can approximate
\begin{equation}\begin{split}
\tilde{T}_{1,s,s} \overset{z\gg 1}{\approx}& -\frac{2}{\sqrt{\pi}} z^{-3/4} e^{-\frac{2}{3} z^{3/2}}  \frac{p_-}{p'_-}, \quad \text{for $ s=-\text{sgn}(\psi'(\phi_+))$}.
\end{split}\end{equation}

For opposite spin quantum numbers $(s=-s')$ the asymptotic expression is
\begin{equation}\begin{split}
\tilde{T}_{1,s,-s} \overset{z\gg 1}{\approx}
 - \frac{1}{4\sqrt{\pi}} z^{-9/4} e^{-\frac{2}{3} z^{3/2}} \frac{p_-}{p'_-} .
\end{split}\end{equation}
With photon polarization $j=2$ we have for identical spin quantum numbers $(s=s')$
\begin{equation}\begin{split}
\tilde{T}_{2,s,s} \overset{z\gg 1}{\approx}
 - \frac{1}{4\sqrt{\pi}} z^{-9/4} e^{-\frac{2}{3} z^{3/2}}  \frac{p_-}{p'_-} 
\end{split}\end{equation}
and for opposite spin quantum numbers $(s=-s')$
\begin{equation}\begin{split}
\tilde{T}_{2,s,-s} \overset{z\gg 1}{\approx}& -\frac{1}{\sqrt{\pi}} z^{-3/4} e^{-\frac{2}{3} z^{3/2}} \frac{p_-}{p'_-} \left[1+s\ \text{sgn}(\psi'(\phi_+))\right]\\
& -\frac{1}{9216 \sqrt{\pi}} z^{-15/4} e^{-\frac{2}{3} z^{3/2}} \frac{p_-}{p'_-} \left[ 3938  +770s\ \text{sgn}(\psi'(\phi_+)) \right] .
\end{split}\end{equation}
Analogously as above, the scaling of the emission probability will be determined by the approximated expression
\begin{equation}
\tilde{T}_{2,s,-s} \overset{z\gg 1}{\approx} -\frac{2}{\sqrt{\pi}} z^{-3/4} e^{-\frac{2}{3} z^{3/2}} \frac{p_-}{p'_-}, \quad \text{for $s=\text{sgn}(\psi'(\phi_+))$}.
\end{equation}

\subsection{Nonlinear Breit-Wheeler pair production}

Now, we pass to the probability of nonlinear Breit-Wheeler pair production including the particle states decay. 

Here, the incoming photon has four-momentum $q^{\mu}=(\omega,\bm{q})$, with polarization quantum number $j=1,2$ and the outgoing positron (electron) has four-momentum $p^{\mu}=(\varepsilon,\bm{p})$ ($p'^{\mu}=(\varepsilon',\bm{p}')$) and spin quantum number $s=\pm 1$ ($s'=\pm 1$). The symbols of quantum numbers of the particles have been chosen in order to exploit the crossing symmetry between the amplitudes of nonlinear Compton scattering and of nonlinear Breit-Wheeler pair production. The expression of the probability was computed in Ref. \cite{Podszus_2021} and it is given by
\begin{equation}
\label{P_NBW}
\begin{split}
P^{(\gamma \to e^- e^+)}_{j,s,s'}=&\int\frac{d^3p}{16\pi^2}\frac{\alpha}{q_-p'_-\varepsilon}\int d\phi_+e^{2\text{Im}\left\{\frac{m}{q_-}\int_{-\infty}^{\phi_+}d\varphi P_j(q,\varphi)+\int_{\phi_+}^{\infty}d\varphi \left[\frac{m}{p'_-}M_{s'}(p',\varphi)+\frac{m}{p_-}M_s(-p,\varphi)\right]\right\}}\\
&\times\int d\phi_- e^{i\frac{m^2}{2p_-}\frac{q_-}{p'_-}\left\{[1+\bm{\pi}_{\perp,p}^2(\phi_+)]\phi_-+\frac{\bm{\mathcal{E}}^2(\phi_+)}{m^2}\frac{\phi_-^3}{12}\right\}} G_{j,s,s'},
\end{split}
\end{equation}
with the trace
\begin{equation}\label{Trace_NBW}
\begin{split}
G_{j,s,s'}=\frac{1}{4}\text{tr}&\left\{\left[1-\frac{\hat{n}[\hat{\mathcal{A}}(\phi_+)+\hat{\mathcal{A}}'(\phi_+)\phi_-/2]}{2p'_-}\right]\hat{\Lambda}_j(q)
\left[1-\frac{\hat{n}[\hat{\mathcal{\mathcal{A}}}(\phi_+)+\hat{\mathcal{A}}'(\phi_+)\phi_-/2]}{2p_-}\right]\right.\\
&\left.\times(\hat{p}-m)(1+s\gamma^5\hat{\zeta}) \left[1+\frac{\hat{n}[\hat{\mathcal{A}}(\phi_+)-\hat{\mathcal{A}}'(\phi_+)\phi_-/2]}{2p_-}\right]
\hat{\Lambda}_j(q)\right.\\
&\left.\times\left[1+\frac{\hat{n}[\hat{\mathcal{A}}(\phi_+)-\hat{\mathcal{A}}'(\phi_+)\phi_-/2]}{2p'_-}\right](\hat{p}'+m)(1+s'\gamma^5\hat{\zeta}')\right\}
\end{split}
\end{equation}
and with
\begin{equation}
\bm{\pi}_{\perp,p}(\phi)=\frac{\bm{p}_{\perp}}{m}-\frac{p_-}{q_-}\frac{\bm{q}_{\perp}}{m}+\frac{\bm{\mathcal{A}}_{\perp}(\phi)}{m}.
\end{equation}
The trace can be simplified and the integrals in the perpendicular positron momentum and $\phi_-$ can be taken analogously as in the case of nonlinear Compton scattering. An important difference is however that the relations for momentum conservation are instead $\bm{p}'_{\perp}=\bm{q}_{\perp} -\bm{p}_{\perp}$ and $p'_-=q_- -p_-$. 

By performing the trace over the matrix contractions (see the appendix for the derivation), the expressions for the two polarization quantum numbers $j=1$ and $j=2$ become
\begin{equation}\begin{split}\label{G_NBW_1}
G_{1,s,s'} &= (1+s s') \bigg[ \frac{1}{2} \frac{q_-^2}{p_-p'_-} \left(m^2 +m^2 \bm{\pi}_{\perp ,p}^2(\phi_+) -\mathcal{A}_0^2 \psi'^2(\phi_+) \frac{\phi_-^2}{4} \right) \\
&\qquad \qquad \quad -2 \left(m^2 \bm{\pi}_{\perp ,p}^2(\phi_+) -\mathcal{A}_0^2 \psi'^2(\phi_+) \frac{\phi_-^2}{4} \right) \bigg]\\
&-i (s+s') \frac{m}{2} \mathcal{A}_{0} \psi'(\phi_+) \phi_- \frac{q_-}{p_-}\left(2-\frac{q_-}{p'_-} \right)\\
&+ \left(2+2ss'- s s' \frac{q_-^2}{p_- p'_-}\right) \left(p_2-\frac{p_-}{q_-} q_2 \right)^2
\end{split}\end{equation}
and 
\begin{equation}\begin{split}\label{G_NBW_2}
G_{2,s,s'} &= (1-s s') \frac{1}{2} \frac{q_-^2}{p_-p'_-} \left( m^2 +m^2 \bm{\pi}_{\perp,p}^2(\phi_+) -\mathcal{A}_0^2 \psi'^2(\phi_+) \frac{\phi_-^2}{4} \right) \\
&-i (s-s') \frac{m}{2} \mathcal{A}_{0} \psi'(\phi_+) \phi_- \frac{q_-^2}{p_-p'_-}\\
&-\left(2+2ss' - s s' \frac{q_-^2}{p_-p'_-}\right) \left(p_2-\frac{p_-}{q_-} q_2 \right)^2 .
\end{split}\end{equation}
Here we have already expressed the pulse shape function in terms of $\bm{\pi}_{\perp ,p}(\phi_+)$. The dependence on $\bm{\pi}_{\perp ,p}(\phi_+)$ can in turn be removed by using the fact that the integral over terms proportional to $\left[1+ \bm{\pi}_{\perp,p}^2(\phi_+) +( {\mathcal{A}^2_0 \psi'^2(\phi_+)}/{m^2}) ({\phi_-^2}/{4})\right]$ vanish when performing the integral in $\phi_-$, due to the properties of the Airy functions. In this way, by adding and subtracting suitable terms, the traces can be written in a manifestly gauge-invariant way.
Furthermore, we transform the integral in the positron momentum into light-cone coordinates by using that $d^3 p= (\varepsilon/p_-) dp_- d^2 p_{\perp} $ and we employ the notation
\begin{equation}
\tilde{G}_{j,s,s'}=-\frac{1}{4\pi^2 m^2} \frac{q_-}{p_- p'_-} \int d\phi_- \int d^2p_{\perp} e^{i\frac{m^2 q_-}{2p_- p'_-}\left\{[1+\bm{\pi}_{\perp,p}^2(\phi_+)]\phi_-+\frac{\bm{\mathcal{E}}^2(\phi_+)}{m^2}\frac{\phi_-^3}{12}\right\}} G_{j,s,s'}.
\end{equation}
The integral over the transverse positron momentum $\bm{p}_{\perp}$ is taken by employing the two Gaussian integrals \cite{NIST_b_2010}
\begin{align}
&\int d^2 p_{\perp} e^{i \frac{m^2 q_-}{2p_- p'_-} \bm{\pi}^2_{\perp,p} (\phi_+) \phi_-} =2\pi i \frac{p_- p'_-}{q_- (\phi_- +i0)} , \\
&\int d^2 p_{\perp} \left(p_2-\frac{p_-}{q_-} q_2 \right)^2 e^{i \frac{m^2 q_-}{2p_- p'_-} \bm{\pi}^2_{\perp,p} (\phi_+) \phi_-} = -2\pi \left[\frac{p_- p'_-}{q_- (\phi_- +i0)}\right]^2,
\end{align}
and the integral in $\phi_-$ results again in Airy functions. In this way, the probability of nonlinear Breit-Wheeler pair production including the damping of particle states finally reads
\begin{equation}\begin{split}\label{P_NBW_final}
P^{(\gamma \to e^- e^+)}_{j,s,s'} = &
- \frac{\alpha m^2}{4 q_-^2} \int_0^{q_-} dp_- \int d\phi_+ \tilde{G}_{j,s,s'}\\
&\times e^{2\text{Im} \left\{\frac{m}{q_-} \int_{-\infty}^{\phi_+} d\varphi P_j(q,\varphi) +\int_{\phi_+}^{\infty} d\varphi \left[ \frac{m}{p'_-} M_{s'}(p',\varphi) +\frac{m}{p_-} M_s(-p,\varphi) \right] \right\} },
\end{split}\end{equation}
with
\begin{equation}
\label{G_1_final}
\begin{split}
\tilde{G}_{1,s,s'}
= & \left[ -(1+ss')- ss' \frac{q_-^2}{2p_-p'_-}\right] \text{Ai}_1(z) \\
&+\left[ -3 (1+ss') +\left(1+\frac{ss'}{2}\right) \frac{q_-^2}{p_-p'_-} \right] \frac{\text{Ai}'(z)}{z}\\
&-(s+s') \left(\frac{q_-}{p_-}- \frac{q_-}{p'_-} \right) \frac{\text{Ai}(z)}{\sqrt{z}} \text{sgn}(\psi'(\phi_+)) 
\end{split}\end{equation}
and 
\begin{equation}
\label{G_2_final}
\begin{split}
\tilde{G}_{2,s,s'}
= &  \left[ -(1+ss')+ ss' \frac{q_-^2}{2p_-p'_-}\right] \text{Ai}_1(z) \\
&+\left[ -(1+ss') +\left(1-\frac{ss'}{2}\right) \frac{q_-^2}{p_-p'_-} \right] \frac{\text{Ai}'(z)}{z}\\
&+(s'-s) \frac{q_-^2}{p_-p'_-} \frac{\text{Ai}(z)}{\sqrt{z}} \text{sgn}(\psi'(\phi_+)) .
\end{split}\end{equation}
Without the exponential damping term the above probability reduces to the result of the spin- and polarization-resolved probability of nonlinear Breit-Wheeler pair production calculated in Ref. \cite{Seipt_2020}. Analogously as in the case of nonlinear Compton scattering, it can be proved analytically that the probability in Eq. (\ref{P_NBW_final}) is always smaller than unity [see the discussion below Eq. (\ref{T_NCS_f2})].

Now, we investigate the asymptotic behavior in the two regions $q_--p_-\ll q_-$ and $p_-\ll q_-$ and for the differential probability of nonlinear Breit-Wheeler pair production
\begin{equation}\label{diff_prob_NBW}
\begin{split}
\frac{\partial P^{(\gamma \to e^- e^+)}_{j,s,s'}}{\partial p_-} = &
- \frac{\alpha m^2}{4 q_-^2} \int d\phi_+ e^{D^{\text{NBW}}_{j,s,s'} } \tilde{G}_{j,s,s'} ,
\end{split}\end{equation}
with
\begin{equation}\label{c_damp_NBW}
\begin{split}
D^{\text{NBW}}_{j,s,s'}= 2\text{Im} \left\{\frac{m}{q_-} \int_{-\infty}^{\phi_+} d\varphi P_j(q,\varphi) +\int_{\phi_+}^{\infty} d\varphi \left[ \frac{m}{p'_-} M_{s'}(p',\varphi) +\frac{m}{p_-} M_s(-p,\varphi) \right] \right\},
\end{split}\end{equation}
which we obtained from Eq. (\ref{P_NBW_final}). According to the optical theorem, the exponent of the damping term is here equal to minus the sum of the total probability of nonlinear Breit-Wheeler pair production between $-\infty$ and $\phi_+$ for the incoming photon with light-cone energy $q_-$ and polarization quantum number $j$ and the total probabilities of nonlinear Compton scattering between $\phi_+$ and $+\infty$ for the outgoing electron with light-cone energy $p'_-$ and spin quantum number $s'$ and for the outgoing positron with light-cone energy $p_-$ and spin quantum number $s$ \cite{Baier_1976_a,Baier_1976_b}, i.e.
\begin{equation}\label{c_damp_NBW_prob}
\begin{split}
D^{\text{NBW}}_{j,s,s'}= -\int_{-\infty}^{\phi_+} d\varphi \frac{\partial P^{\text{NBW}}_{j,q}}{\partial \varphi} -\int_{\phi_+}^{\infty} d\varphi \left( \frac{\partial P^{\text{NC}}_{s',p'}}{\partial \varphi} +\frac{\partial P^{\text{NC}}_{s,p}}{\partial \varphi} \right).
\end{split}\end{equation}

\subsubsection{\textbf{Asymptotic expression for $q_--p_- \ll q_-$}}
First, we consider the asymptotic region $p'_-= q_--p_-\ll q_-$. Thus, by assuming that the quantum nonlinearity parameter $\kappa_q(\varphi)$ of the photon is fixed, the absolute value of the quantum nonlinearity parameter of the electron $\chi_{p'} (\varphi)=(p'_-/q_-)\kappa_q(\varphi)$ is much smaller than unity (if $\kappa_q(\varphi)$ is larger than unity, the ratio $p'_-/q_-$ is assumed to be sufficiently small that $|\chi_{p'} (\varphi)|\ll 1$) and we use for the damping term in Eq. (\ref{c_damp_NBW_prob}) the corresponding asymptotic expression for the probability of nonlinear Compton scattering, which is independent of $p'$ and given in Eq. (\ref{NCS_chi_small}). The exponent of the damping term becomes (we use the fact that $\chi_p(\varphi)=(p_-/q_-)\kappa_q(\varphi)$)
\begin{equation}\begin{split}
D^{\text{NBW}}_{j,s,s'}
\overset{p'_- \ll q_-}{\approx} \frac{2m}{q_-} \text{Im} \left[ \int_{-\infty}^{\phi_+} d\varphi P_j(q,\varphi) + \int_{\phi_+}^{\infty} d\varphi M_s(-q,\varphi) \right] - \int_{\phi_+}^{\infty} d\varphi \frac{5}{\sqrt{3}} \frac{\alpha m^2 |\kappa_q(\varphi)|}{q_-}.
\end{split}\end{equation}
Concerning the function $\tilde{G}_{j,s,s'}$, we can expand the Airy functions for $z=\left( \frac{q_-}{p'_- \chi_p(\phi_+)}\right)^{2/3} \approx \left( \frac{q_-}{p'_- \kappa_q(\phi_+)}\right)^{2/3} \gg 1$.
In the case of photon polarization $j=1$ and identical spin quantum numbers $(s=s')$, we obtain the expression
\begin{equation}\begin{split}\label{G_z_infty_a}
\tilde{G}_{1,s,s} \overset{z \gg 1}{\approx}& -\frac{1}{\sqrt{\pi}} z^{-3/4} e^{-\frac{2}{3} z^{3/2}} \left[ \frac{q_-}{p'_-} \left(1- s\ \text{sgn}(\psi'(\phi_+)) \right) +2 \frac{p'_-}{q_-} s\ \text{sgn}(\psi'(\phi_+)) \right] \\
& + \frac{1}{96\sqrt{\pi}} z^{-9/4} e^{-\frac{2}{3} z^{3/2}} \left[ 124 +20 s\ \text{sgn}(\psi'(\phi_+)) \right]\\
& -\frac{1}{9216 \sqrt{\pi}} z^{-15/4} e^{-\frac{2}{3} z^{3/2}} \frac{q_-}{p'_-} \left[ 3938 -770 s\ \text{sgn}(\psi'(\phi_+)) \right].
\end{split}\end{equation}
As in the case of nonlinear Compton scattering, the scaling of the probability is determined by the case $s=-\text{sgn}(\psi'(\phi_+))$, where
\begin{equation}\begin{split}
\tilde{G}_{1,s,s} \overset{z  \gg 1}{\approx}& -\frac{2}{\sqrt{\pi}} z^{-3/4} e^{-\frac{2}{3} z^{3/2}} \frac{q_-}{p'_-} .
\end{split}\end{equation}
For opposite spin quantum numbers $(s=-s')$ we have
\begin{equation}\begin{split}\label{G_z_infty_b}
\tilde{G}_{1,s,-s} \overset{z  \gg 1}{\approx}
 - \frac{1}{4\sqrt{\pi}} z^{-9/4} e^{-\frac{2}{3} z^{3/2}} \frac{q_-}{p'_-} .
\end{split}\end{equation}
With photon polarization $j=2$ the asymptotic expansion is for identical spin quantum numbers $(s=s')$
\begin{equation}\begin{split}\label{G_z_infty_2a}
\tilde{G}_{2,s,s} \overset{z  \gg 1}{\approx}
 - \frac{1}{4\sqrt{\pi}} z^{-9/4} e^{-\frac{2}{3} z^{3/2}}  \frac{q_-}{p'_-} 
\end{split}\end{equation}
and for opposite spin quantum numbers $(s=-s')$
\begin{equation}\begin{split}\label{G_z_infty_2b}
\tilde{G}_{2,s,-s} \overset{z \gg 1}{\approx}& - \frac{1}{\sqrt{\pi}} z^{-3/4} e^{-\frac{2}{3} z^{3/2}}  \frac{q_-}{p'_-} \left[ 1 + s\ \text{sgn}(\psi'(\phi_+)) \right] \\
& -\frac{1}{9216 \sqrt{\pi}} z^{-15/4} e^{-\frac{2}{3} z^{3/2}} \frac{q_-}{p'_-} \left[ 3938 +770 s\ \text{sgn}(\psi'(\phi_+)) \right]  .
\end{split}\end{equation}
The scaling of the probability of nonlinear Breit-Wheeler pair production is determined by the case $s=\text{sgn}(\psi'(\phi_+))$, i.e.,
\begin{equation}\begin{split}\label{G_z_infty_2c}
\tilde{G}_{2,s,-s} \overset{z \gg 1}{\approx}& - \frac{2}{\sqrt{\pi}} z^{-3/4} e^{-\frac{2}{3} z^{3/2}}  \frac{q_-}{p'_-} .
\end{split}\end{equation}

\subsubsection{\textbf{Asymptotic expression for $p_- \ll q_-$}}

Due to the symmetry of the probability of pair production under the exchanges $p_-\leftrightarrow p'_-$,  $s\leftrightarrow s'$, and $\psi'(\phi_+)\leftrightarrow -\psi'(\phi_+)$ [see Eqs. (\ref{P_NBW_final})-(\ref{G_2_final})], the asymptotic expressions of the differential probability of nonlinear Breit-Wheeler pair production in the region $p_-\ll q_-$ can be easily obtained from those derived in the region $p'_-\ll q_-$ via the corresponding substitution rules.

\subsection{Arbitrary spin and polarization basis}

So far, both here and in Ref. \cite{Podszus_2021}, the calculations were performed by employing a special direction for the electron or positron spin and for the photon polarization. This choice of the spin and polarization four-vectors has the advantage that the mass and polarization operator become diagonal. Consequently, the equations including the damping of the states can be solved in a relatively straightforward way. However, if the final probabilities are summed over a spin and/or a polarization quantum number, the results should be in the end independent of the choice of the corresponding spin and polarization basis, which we would like to prove explicitly below. 

In the case of the spin state of an electron the two free, positive-energy spinors $u_1(p)$ and $u_{-1}(p)$ form a spin basis \cite{Landau_b_4_1982}. They are normalized as $u^{\dag}_s(p)u_{s'}(p)=2\varepsilon \delta_{ss'}$, with $s,s'=\pm 1$ and fulfill the relation $\gamma^5 \hat{\zeta} u_s (p)=s u_s(p)$ (for the positron the spin basis is formed by the two free negative-energy spinors $v_1(p)$ and $v_{-1}(p)$, normalized as $v^{\dag}_s(p)v_{s'}(p)=2\varepsilon \delta_{ss'}$) \cite{Landau_b_4_1982}. A spinor corresponding to an arbitrary spin direction, indicated here as $u_{+}(p)$, can be expressed by a linear combination of the two basis spinors $u_1(p)$ and $u_{-1}(p)$ as
\begin{equation}\label{u_+}
u_{+}(p)=\beta_1 u_1(p) +\beta_{-1} u_{-1}(p) ,
\end{equation}
where $\beta_1 ,\beta_{-1}$ are two complex numbers such that $|\beta_1 |^2 +|\beta_{-1} |^2=1$. The two coefficients $\beta_1$ and $\beta_{-1}$ are related to the polar angle $\theta$ and the azimuthal angle $\varphi$ between the spin vector $\bm{\zeta}$ and the new spin axis \cite{Landau_b_4_1982}. The spinor $u_+(p)$ in Eq. (\ref{u_+}) forms a basis together with the spinor
\begin{equation}
u_{-}(p)=\beta^*_{-1} u_1(p) -\beta^*_{1} u_{-1}(p) ,
\end{equation}
which is perpendicular to $u_+ (p)$. Now, in Ref. \cite{Podszus_2021} the probabilities were calculated by using the exact electron, positron and photon states, which were obtained by solving the corresponding Schwinger-Dyson equations. For the electron out-state $\Psi_{e}^{(\text{out})}(x)$ the Schwinger-Dyson equation is $\{\gamma^{\mu}[i\partial_{\mu}-\mathcal{A}_{\mu}(\phi)]-m\}\Psi_{e}^{(\text{out})}(x)=\int d^4y\, \bar{M}_L(y,x)\Psi_e^{(\text{out})}(y)$, with $\bar{M}_L(y,x)=\gamma_0 M_L^{\dagger} (y,x) \gamma_0$ and $M_L (y,x)$ being the mass operator in a plane wave. This equation is linear in the spin basis $u_1(p)$ and $u_{-1}(p)$, such that we can decompose an arbitrary electron state in terms of the two states $\Psi_{e,1}^{(\text{out})}(x)$ and $\Psi_{e,-1}^{(\text{out})}(x)$, which are solutions of the Schwinger-Dyson equation constructed via the spinors $u_1(p)$ and $u_{-1}(p)$, respectively.
The electron out-state $\Psi_{e,b}^{(\text{out})} (x)$ of an arbitrary spin direction $b=\{+,-\}$, given by the solution of the Schwinger-Dyson equation for the spinor $u_b (p)$, can be expressed then by the linear combination 
\begin{equation}\begin{split}
\Psi^{(\text{out})}_{e,+}(x)=&\beta_1 \Psi^{(\text{out})}_{e,1}(x) +\beta_{-1} \Psi^{(\text{out})}_{e,-1}(x),\\
\Psi^{(\text{out})}_{e,-}(x)=&\beta^*_{-1} \Psi^{(\text{out})}_{e,1}(x) -\beta^*_{1} \Psi^{(\text{out})}_{e,-1}(x).
\end{split}\end{equation}
The physical meaning of the states with the above choice of $\beta_1$ and $\beta_{-1}$ is that at $x^0\to\infty$ in the rest frame of the electron the spin axis points along the chosen axis and fulfill the above-mentioned relations. Considering now the process of nonlinear Compton scattering, the $S$-matrix element for an incoming electron of spin quantum number $s$, an outgoing photon of polarization $j$, and an outgoing electron with spin quantum number $b$ is proportional to
\begin{equation}
S^{(e^-\to e^-\gamma)}_{s,j,+}=\beta^*_1 S^{(e^-\to e^-\gamma)}_{s,j,1} +\beta^*_{-1} S^{(e^-\to e^-\gamma)}_{s,j,-1}
\end{equation}
and
\begin{equation}
S^{(e^-\to e^-\gamma)}_{s,j,-}=\beta_{-1} S^{(e^-\to e^-\gamma)}_{s,j,1} -\beta_{1} S^{(e^-\to e^-\gamma)}_{s,j,-1}.
\end{equation}
By using these $S$-matrix elements, we then easily obtain the probability of nonlinear Compton scattering for the two spin quantum numbers $+$ and $-$ of the final electron as
\begin{equation}\begin{split}\label{P_+}
P^{(e^-\to e^-\gamma)}_{s,j,+} &=\int \frac{d^3q}{(2\pi)^3} \frac{d^3p'}{(2\pi)^3} |S^{(e^-\to e^-\gamma)}_{s,j,+}|^2 \\
&=\int \frac{d^3q}{(2\pi)^3} \frac{d^3p'}{(2\pi)^3} \Big[ |\beta_1|^2 |S^{(e^-\to e^-\gamma)}_{s,j,1}|^2 +|\beta_{-1}|^2 |S^{(e^-\to e^-\gamma)}_{s,j,-1}|^2\\
& \qquad \qquad \ \ \ + \beta_1 \beta^*_{-1} S^{(e^-\to e^-\gamma)\,*}_{s,j,1} S^{(e^-\to e^-\gamma)}_{s,j,-1} + \beta^*_1 \beta_{-1} S^{(e^-\to e^-\gamma)}_{s,j,1} S^{(e^-\to e^-\gamma)\,*}_{s,j,-1} \Big]
\end{split}\end{equation}
and
\begin{equation}\begin{split}\label{P_-}
P^{(e^-\to e^-\gamma)}_{s,j,-} &=\int \frac{d^3q}{(2\pi)^3} \frac{d^3p'}{(2\pi)^3} |S^{(e^-\to e^-\gamma)}_{s,j,-}|^2 \\
&=\int \frac{d^3q}{(2\pi)^3} \frac{d^3p'}{(2\pi)^3} \Big[ |\beta_{-1}|^2 |S^{(e^-\to e^-\gamma)}_{s,j,1}|^2 +|\beta_{1}|^2 |S^{(e^-\to e^-\gamma)}_{s,j,-1}|^2 \\
& \qquad \qquad \ \ \ - \beta_1 \beta^*_{-1} S^{(e^-\to e^-\gamma)\,*}_{s,j,1} S^{(e^-\to e^-\gamma)}_{s,j,-1} - \beta^*_1 \beta_{-1} S^{(e^-\to e^-\gamma)}_{s,j,1} S^{(e^-\to e^-\gamma)\,*}_{s,j,-1} \Big] .
\end{split}\end{equation}
Taking now the sum of the probabilities for both spin states $b=\{ +,-\}$, i.e., taking the sum of Eqs. (\ref{P_+}) and (\ref{P_-}), we obtain
\begin{equation}\begin{split}
\sum_{b=\{ +,-\} } P^{(e^-\to e^-\gamma)}_{s,j,b} &=\int \frac{d^3q}{(2\pi)^3} \frac{d^3p'}{(2\pi)^3} \Big[ |S^{(e^-\to e^-\gamma)}_{s,j,+}|^2 + |S^{(e^-\to e^-\gamma)}_{s,j,-}|^2 \Big] \\
&= \int \frac{d^3q}{(2\pi)^3} \frac{d^3p'}{(2\pi)^3} \Big[ |S^{(e^-\to e^-\gamma)}_{s,j,1}|^2 + |S^{(e^-\to e^-\gamma)}_{s,j,-1}|^2 \Big] = \sum_{s'=\{ 1,-1\} } P^{(e^-\to e^-\gamma)}_{s,j,s'},
\end{split}\end{equation}
which is identical to the analogous result obtained for the original spin axis. Hence, the probability summed over a spin quantum number is independent of the used quantization axis. It can easily be shown that the same holds for electron in-states and for the positron in- and out-states. 

In order to describe the photon polarization states, we have chosen the two four-vectors $\Lambda^{\mu}_1(q)$ and $\Lambda^{\mu}_{2}(q)$. An arbitrary polarization basis is given by the two four-vectors 
\begin{equation}\begin{split}
\Lambda^{\mu}_{+}(q)=b_1 \Lambda^{\mu}_1(q) +b_{2} \Lambda^{\mu}_{2}(q) \\
\Lambda^{\mu}_{-}(q)=b_{2} \Lambda^{\mu}_1(q) -b_{1} \Lambda^{\mu}_{2}(q) 
\end{split}\end{equation}
with $b_1,b_2$ being two real numbers such that $b_1^2 +b_2^2 =1$. Analogously to the case of the spin, one can show that also in this case the probability summed over the polarization indexes does not depend on the polarization basis.

\section{Numerical calculations}

\subsection{Methods}

The results in the following section have been produced using two types of integration: 1) a purely numerical one in order to obtain probability densities, fully differential with respect to an outgoing particle's momentum and 2) a quadrature which takes advantage of the fact that in lightfront coordinates the transverse momenta yield analytical Gaussian integrals, thus leaving just one longitudinal momentum integral to be done numerically in order to derive the total probability of the process.

We successfully compared the two methods against each other, by integrating the results of method 1 with respect to the transverse momenta and leaving aside the longitudinal integral in method 2. In fact, since the longitudinal momentum is practically proportional to the energy for the ultrarelativistic particles we consider, the latter is a good method of obtaining the energy spectrum.
In both methods, before proceeding to perform the $\phi_+$ integral in Eqs. (\ref{P_NCS}) and (\ref{P_NBW}) we expressed the integrals over $\phi_-$ through Airy functions and stored them in an interpolation table in logarithmic form. Then, we also produced interpolation tables that store, for different spin and polarization numbers and longitudinal momentum values, the mass and polarization integrals in Eqs. (\ref{M_s}) and (\ref{P_j}), respectively. These were then used in computing differential probabilities and total ones. All the numerical quadratures were performed using adaptive Gauss-Konrod rules.

An important step for the production of fully differential spectra was to divide the $\phi_+$ integration domain into some well-chosen subintervals so that the adaptive integrator did not miss any relevant region. For large values of $\xi_0$ the integrand has some very narrow peaks apart from which it is almost vanishing and, unless the integrator is led to those peaks, it would stop short of finding them, yielding a negligible result for the whole integral. To find the peaks we first had to identify and store the intervals of monotonicity of the vector potential and then identify the possible peaks in each such interval using a nonlinear equation solver. As the positions of these sharp peaks change with the transverse momenta, a prior analytical (sharp) Gaussian integration with respect to the latter yielded a $\phi_+$ integral that is easy to handle without any subdivision of intervals.
For the Compton case the longitudinal momentum adaptive quadrature has been made faster through a deformation of its variable, to reduce the number of subdivisions needed to get a good precision at very small $q_-$, while keeping in mind, though, that the behavior of the LCFA result is different from the exact one in this limit.

\subsection{Results}

In this section we present plots based on numerical implementations of the analytical results obtained above. We use the linearly-polarized (along the $x$ direction) plane-wave laser pulse with Gaussian envelope described by the vector potential
\begin{equation}\begin{split}\label{fieldDef}
{\bm A}(\phi)=A_0 e^{-(\phi/\tau)^2} \sin(\omega_0 \phi) {\bm a}_1.
\end{split}\end{equation}
All plots have been made for a carrier angular frequency $\omega_0$ corresponding to $1.55\ \text{eV}$ in our units, which also corresponds to a wavelength of $0.8\ \mu\text{m}$. The parameter $\tau$ describes the length of the pulse and we have chosen the two values $\tau=5$ fs and $\tau=20$ fs. Within the full width at half maximum (FWHM) of the intensity the pulse contains about 2.2 cycles for $\tau=5$ fs and 8.8 cycles for $\tau=20$ fs. Since our analytical results are valid within the LCFA, we consider below amplitudes of the vector potential corresponding to $\xi_0=5$ and we also restrict the parameters $\eta_0= \chi_0 / \xi_0=(k_0p)/m^2$ for nonlinear Compton scattering and $\rho_0= \kappa_0 / \xi_0=(k_0q)/m^2$ for nonlinear Breit-Wheeler pair production, with $k^{\mu}_0=\omega_0n^{\mu}$ to values not exceeding (approximately) unity, by setting an upper bound of $100\ \text{GeV}$ for the incoming particle's energy. For larger values of $\xi_0$ one can further relax the condition on $\eta_0$ and $\rho_0$, allowing them to take even larger values.

We present both probability densities, differential with respect to the momentum of one of the outgoing particles (which determines the momentum of the other outgoing particle through the conservation laws) and total probabilities. Concerning the probabilities integrated over the unconstrained momentum of one of the final particles [see Eqs. (\ref{P_NCS_f})-(\ref{T_NCS_f2}) and Eqs. (\ref{P_NBW_final})-(\ref{G_2_final})], we observe the following. The meaning of the spin and polarization quantum numbers implicitly depends on the momenta of the particles, because the spin and polarization four-vectors depend on the particles' momenta. Thus, in general, once one integrates over the unconstrained momentum of one of the final particles, the physical meaning of the discrete quantum numbers of the final particles is unclear. However, we will always consider head-on collisions with the incoming particle having an energy much larger than $m\xi_0$, such that the angular spread of the produced particles is small. In this case, one can then conclude that the spin and polarization linearly-independent directions approximately correspond to the directions of the electric and magnetic field of the (linearly-polarized) background plane wave in the laboratory frame. Within this approximated framework one can then investigate, for instance, the occurrence of electron spin flip for nonlinear Compton scattering and the distinction between same-spin and different-spin states of the pair produced in nonlinear Breit-Wheeler process. It is interesting to discriminate between such cases, for their probabilities can be very different.

It is convenient to introduce the notation $P^{(e^- \to e^- \gamma)}_s=\sum_{j,s'} P^{(e^- \to e^- \gamma)}_{j,s,s'}$ and $P^{(\gamma \to e^- e^+)}_j=\sum_{s,s'} P^{(\gamma \to e^- e^+)}_{j,s,s'}$ for nonlinear Compton scattering and nonlinear Breit-Wheeler pair production, respectively. 

\begin{figure}
\begin{center}
\includegraphics[width=0.9\columnwidth]{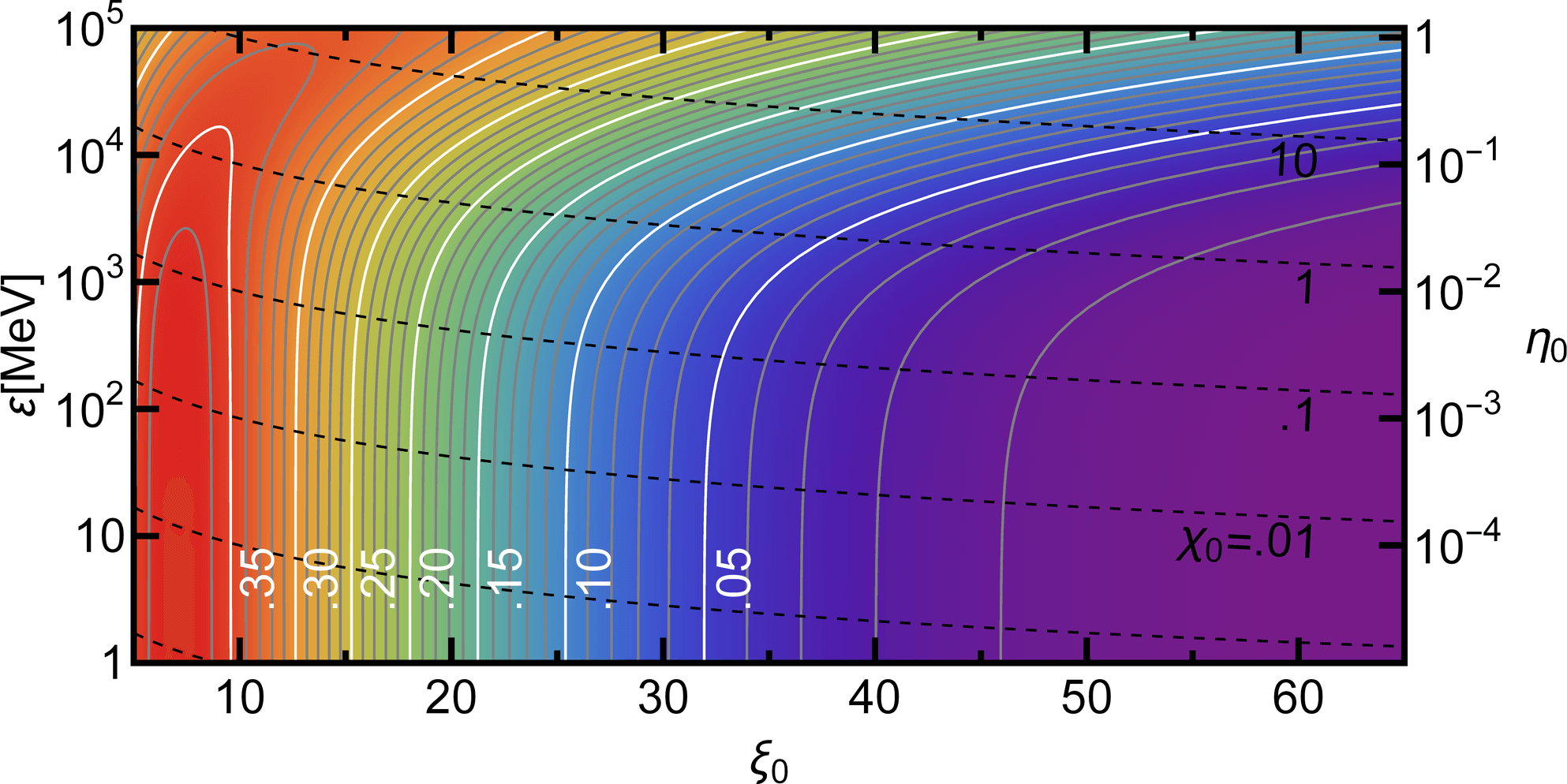}
\includegraphics[width=0.9\columnwidth]{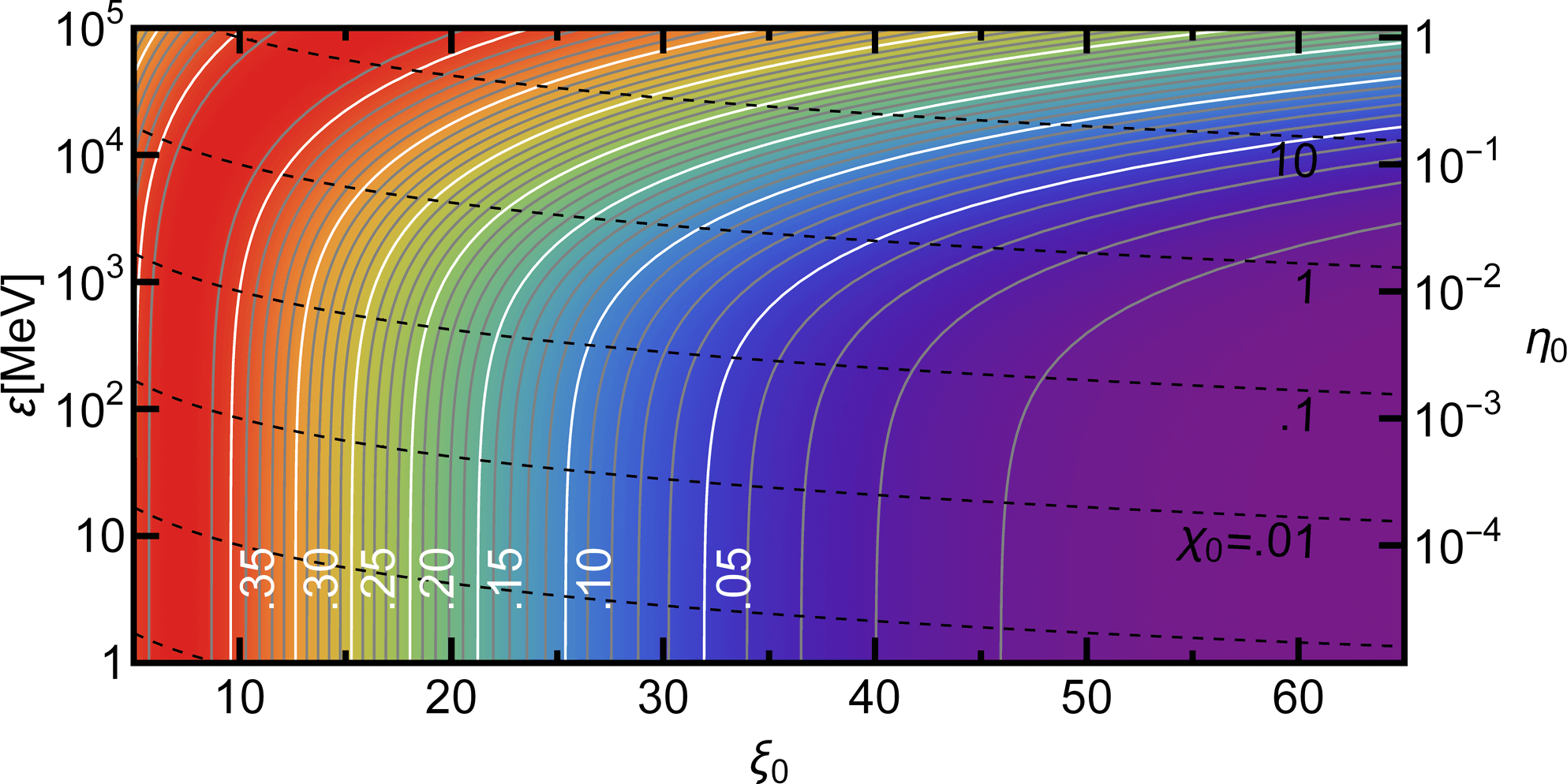}
\caption{Nonlinear Compton scattering total probability $P^{(e^- \to e^- \gamma)}_s$ including the decay of the wave functions (top panel), as compared to the result obtained from a Poissonian distribution whose average photon number is the total ``undamped emission probability'' $P^{\text{NC}}_{s,p}$ (bottom panel). The pulse length corresponds to $\tau=5\;\text{fs}$ and the initial electron spin corresponds to $s=1$.}
\label{Figure_2}
\end{center}
\end{figure}

\begin{figure}
\begin{center}
\includegraphics[width=0.9\columnwidth]{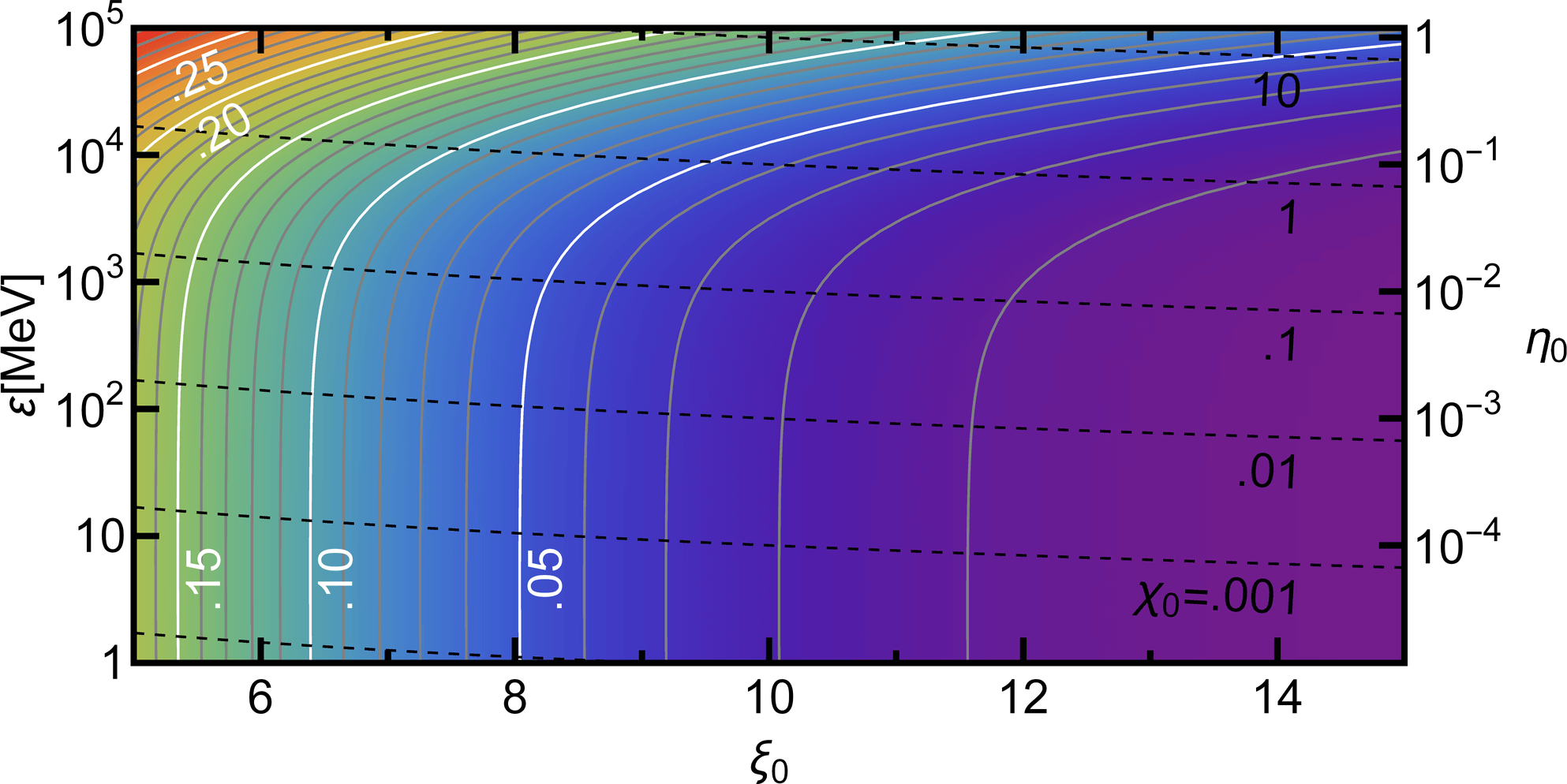}
\includegraphics[width=0.9\columnwidth]{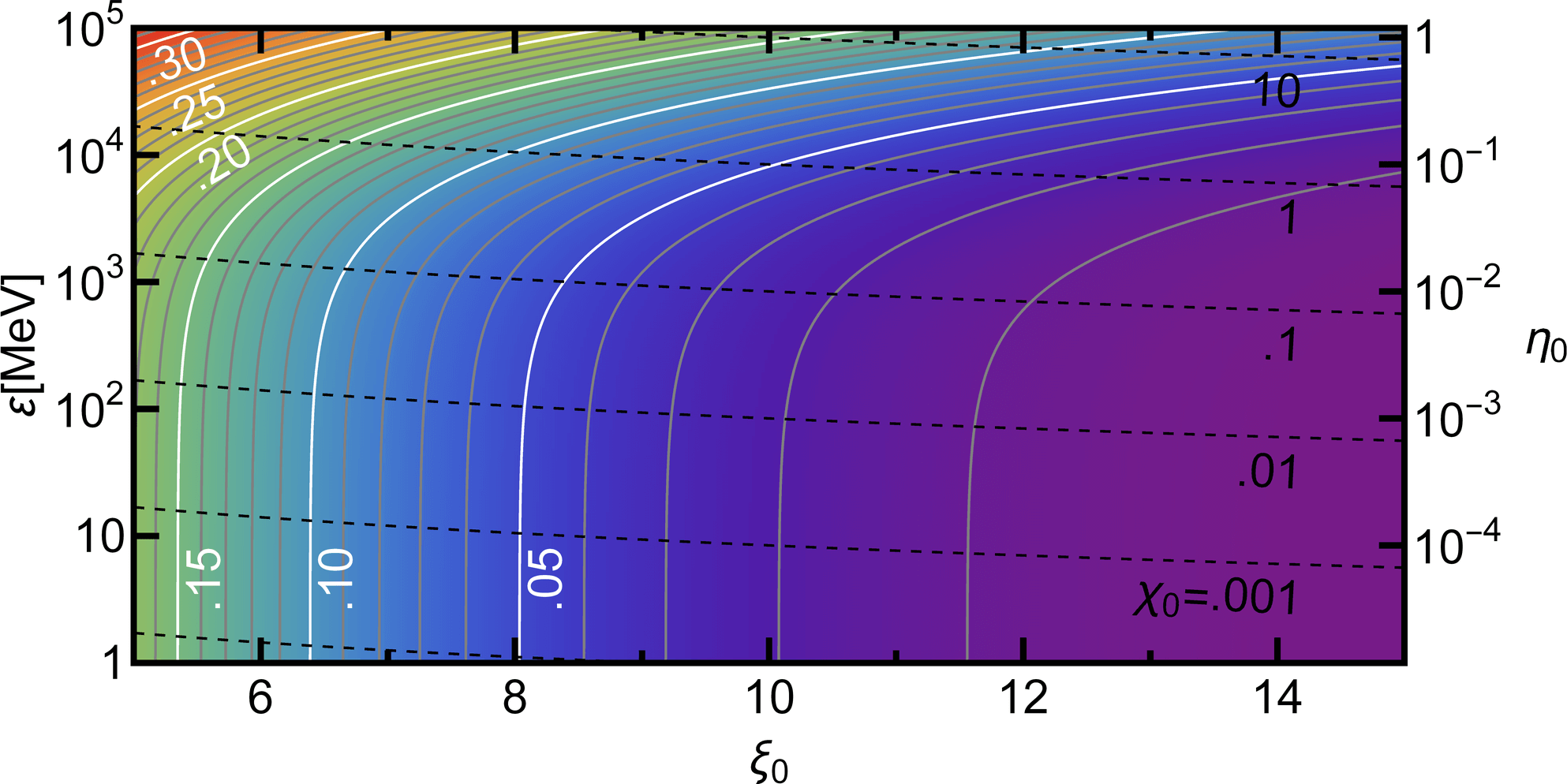}
\caption{Same as in Fig. \ref{Figure_2}, but for $\tau=20\;\text{fs}$.}
\label{Figure_3}
\end{center}
\end{figure}

In Figs. \ref{Figure_2} and \ref{Figure_3} we show the total probability $P^{(e^- \to e^- \gamma)}_s$ of nonlinear Compton scattering (top panels) and the one-event value of the Poisson distribution corresponding to the average $P^{\text{NC}}_{s,p}$ given by the first-order probability of nonlinear Compton scattering, computed by ignoring the decay of particles states, i.e. [see also the discussion below Eq. (\ref{T_NCS_f2})], $P^{\text{NC}}_{s,p} =
- ({\alpha m^2})/({4 p_-^2}) \sum_{j,s'} \int_0^{p_-} dq_- \int d\phi_+\ \tilde{T}_{j,s,s'}$ (bottom panels). For sufficiently small values of $\eta_0$, such that $\chi_0 \ll 1$, the photon recoil is negligible and we expect that the emissions are independent of each other. In Ref. \cite{Glauber_1951} it was shown that in this classical limit the probability of emitting an arbitrary number of photons follows a Poissonian distribution. Indeed, we observe that at such low values of $\chi_0$ the Poissonian distribution well approximates the full QED results [see also the discussion below Eq. (\ref{damp_NCS_class})]. As $\chi_0$ increases, important differences start to be seen.

Two pulse length parameters $\tau$ have been considered: $\tau=5$ fs (Fig. \ref{Figure_2}) and $\tau=20$ fs (Fig. \ref{Figure_3}). In the first case the probability increases as $\xi_0$ increases, reaches a maximum at $\xi_0$ smaller than about 10, and then decreases. In the classical regime of low values of $\chi_0$, the maximum value is $e^{-1}$, as predicted by the Poissonian $P^{\text{NC}}_{s,p}e^{-P^{\text{NC}}_{s,p}}$. For the longer pulse used in Fig. \ref{Figure_3}, the whole plot is in the region where the probability decays with $\xi_0$. Indeed, the decay of the states is now stronger and essentially any increase of $\xi_0$ reduces the single-photon emission probability, as multiple photon emissions become favored over the single-photon emission. As mentioned in the introduction, the decay of the particle states becomes significant if the quantity $\alpha \xi_0 \Phi_L$ is of the order of unity or larger. This corresponds to values of $\xi_0\gtrsim 9.9$ and $\xi_0\gtrsim 2.5$ for a pulse with $\tau= 5\ \text{fs}$ (Fig. \ref{Figure_2}) and $\tau= 20\ \text{fs}$ (Fig. \ref{Figure_3}), respectively. Here, we estimated the total phase duration by the FWHM of the intensity, i.e., we set $\Phi_L = \sqrt{2 \ln{2}}\ \omega_0 \tau$. We see that for larger values of $\tau$ the damping effect becomes significant already at smaller values of $\xi_0$.

\begin{figure}
\begin{center}
\includegraphics[width=1\columnwidth]{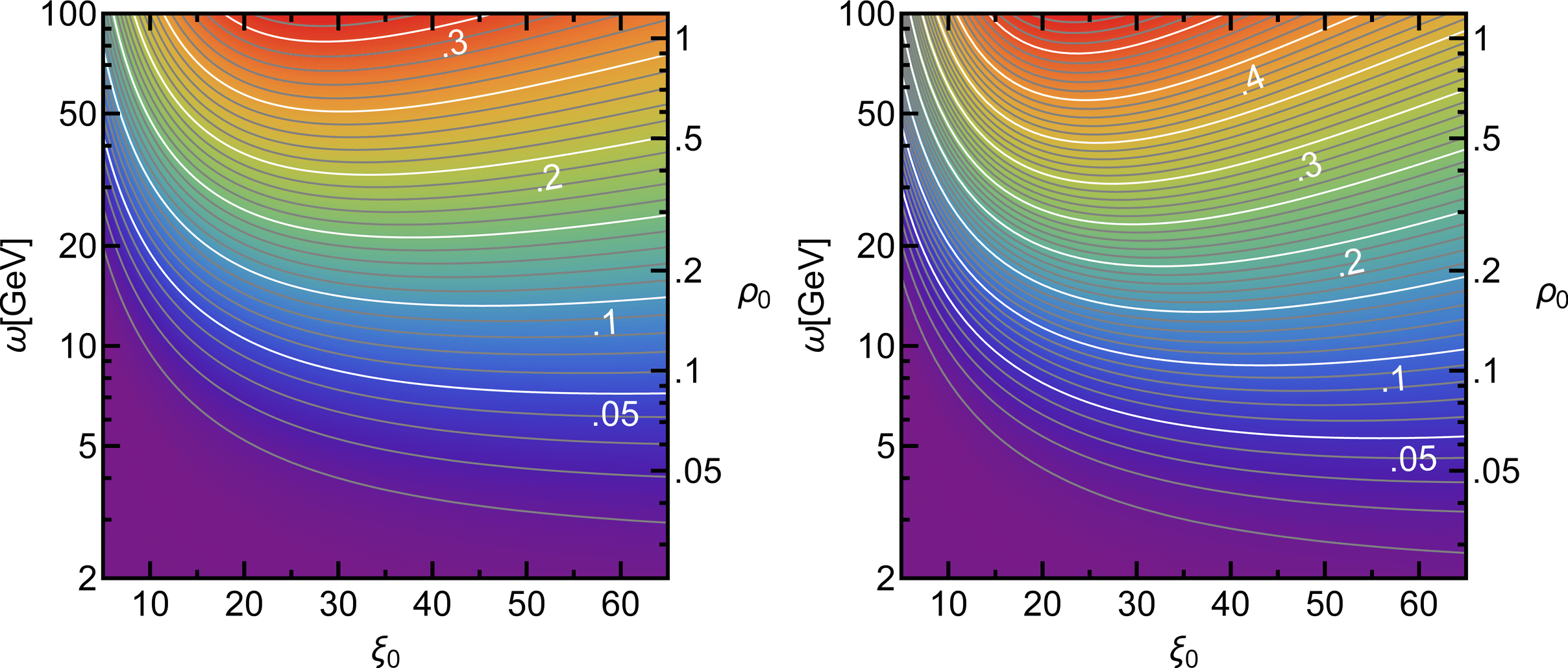}
\caption{The probability $P^{(\gamma \to e^- e^+)}_j$ of nonlinear Breit-Wheeler pair production in a short, $\tau=5\;\text{fs}$ pulse, by a photon with polarization quantum number $j=1$ (left panel) and $j=2$ (right panel).}
\label{Figure_4}
\end{center}
\end{figure}

\begin{figure}
\begin{center}
\includegraphics[width=1\columnwidth]{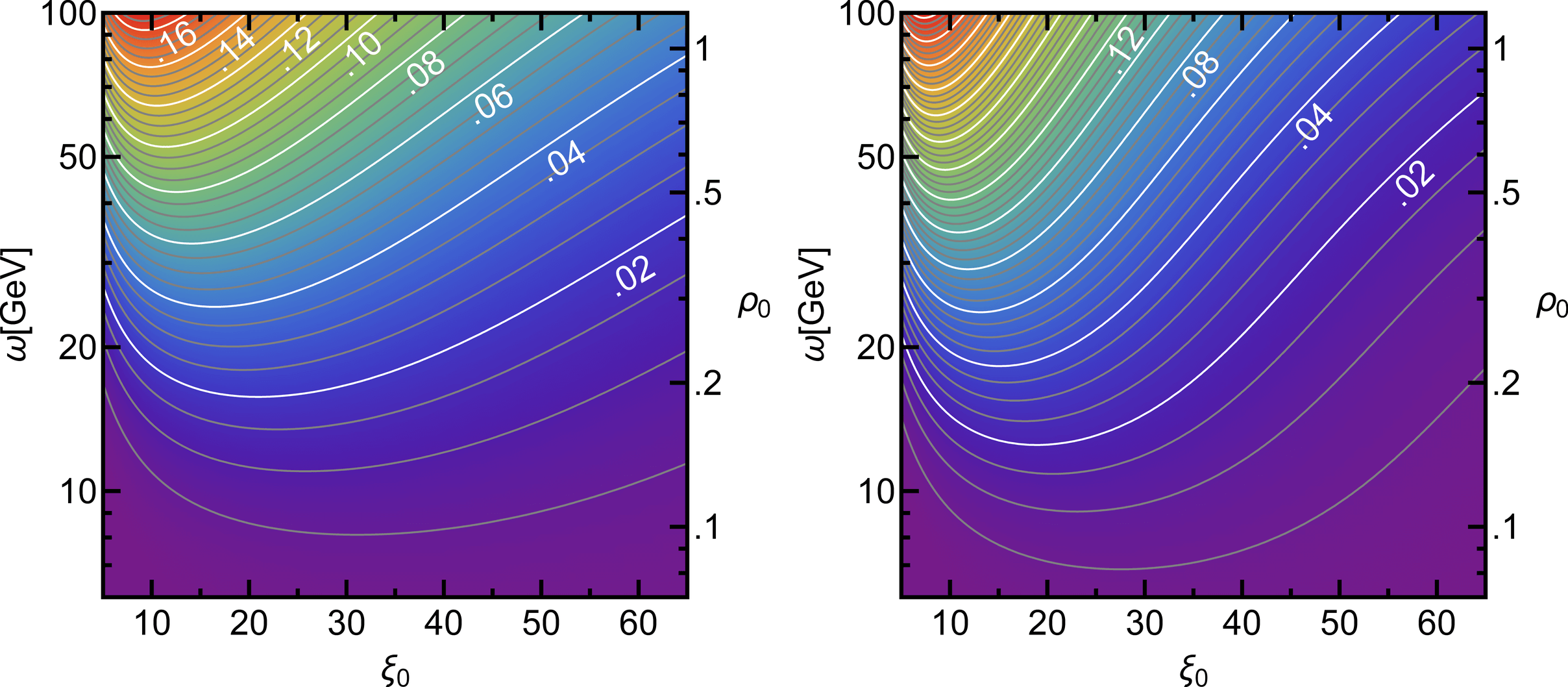}
\caption{Same as in Fig. \ref{Figure_4}, but for $\tau=20\;\text{fs}$.}
\label{Figure_5}
\end{center}
\end{figure}

For the probability of nonlinear Breit-Wheeler pair production we show plots similar to the aforementioned ones, in Figs. \ref{Figure_4} (for a pulse duration corresponding to $\tau=5$ fs) and \ref{Figure_5} (for a pulse duration corresponding to $\tau=20$ fs). A logarithmic scale is used for the energy of the incoming particle in both cases, but in the present case we started from higher values than for nonlinear Compton scattering, due to the exponential suppression of the process that occurs at low $\kappa_0=\rho_0 \xi_0$. In both cases the probability shows a maximum in $\xi_0$ (see also Ref. \cite{Mercuri_Baron_2021}). This feature is observable in the plots only at high incoming photon energies for the shorter pulse, but also at lower incoming photon energies for the longer pulse, indicating that the size of the effect also depends on the damping of the states.

\begin{figure}
\begin{center}
\includegraphics[width=1\columnwidth]{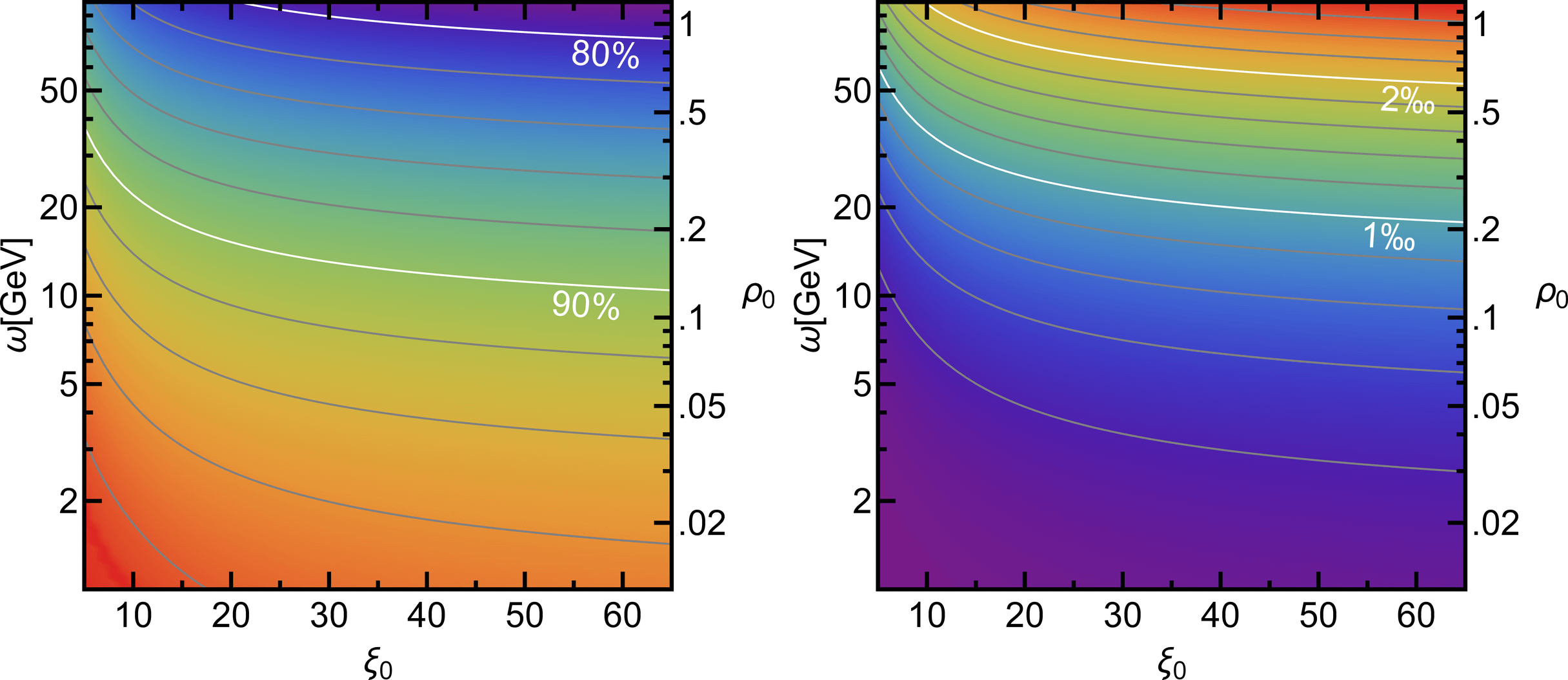}
\caption{Fraction of the total probability of nonlinear Breit-Wheeler pair production corresponding to same-spin pairs, $(\sum_{s} P^{(\gamma \to e^- e^+)}_{j,s,s})/P^{(\gamma \to e^- e^+)}_{j}$ for $j=1$ (left panel) and for $j=2$ (right panel). In both cases, the pulse duration corresponds to $\tau=20\;\text{fs}$.}
\label{Figure_10}
\end{center}
\end{figure}

It can be seen from Fig. \ref{Figure_10} that, as $\rho_0$ increases toward unity, the probability for $(j=1, s=s')$ reduces to around $80\%$ of the total nonlinear Breit-Wheeler pair production probability (implying that for $(j=1, s=-s')$ it grows to around $20\%$), whereas the probability for $(j=2, s=s')$ stays smaller than $3\permil$. Analogous behaviors can be observed in nonlinear Compton scattering, according to the crossing symmetry existing between the two processes.

\begin{figure}
\begin{center}
\includegraphics[width=1\columnwidth]{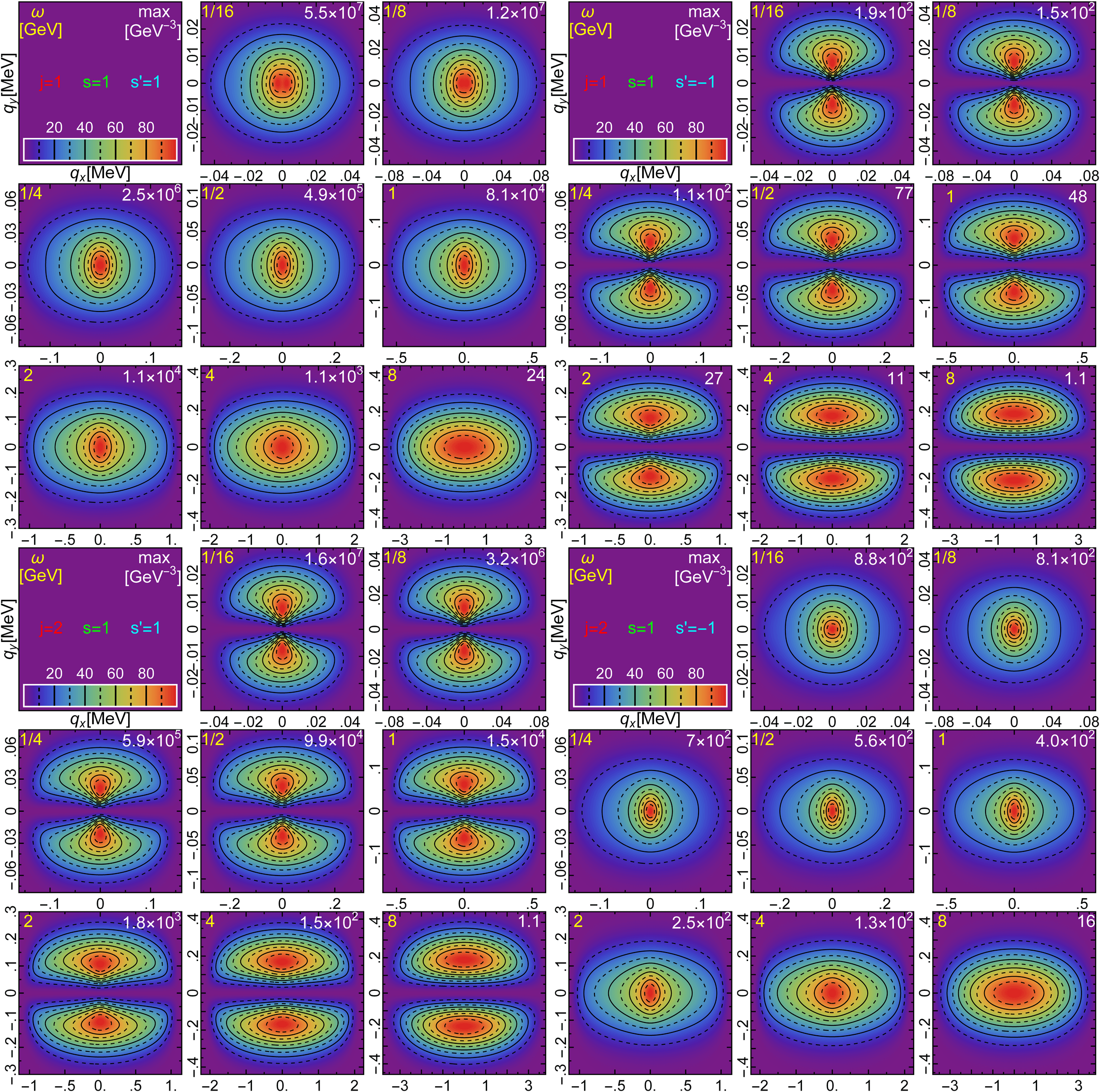}
\caption{Sections through the nonlinear Compton scattering probability distribution in GeV$^{-3}$, for $\tau=20\;\text{fs}$, $\varepsilon=10\;\text{GeV}$, $\xi_0=10$, and incoming electron spin quantum number $s=1$. The color levels indicate percentages of the maximum reached by the distribution in that section. The approximate value of that maximum is shown at the upper right corner of each subplot.}
\label{Figure_6}
\end{center}
\end{figure}

\begin{figure}
\begin{center}
\includegraphics[width=1\columnwidth]{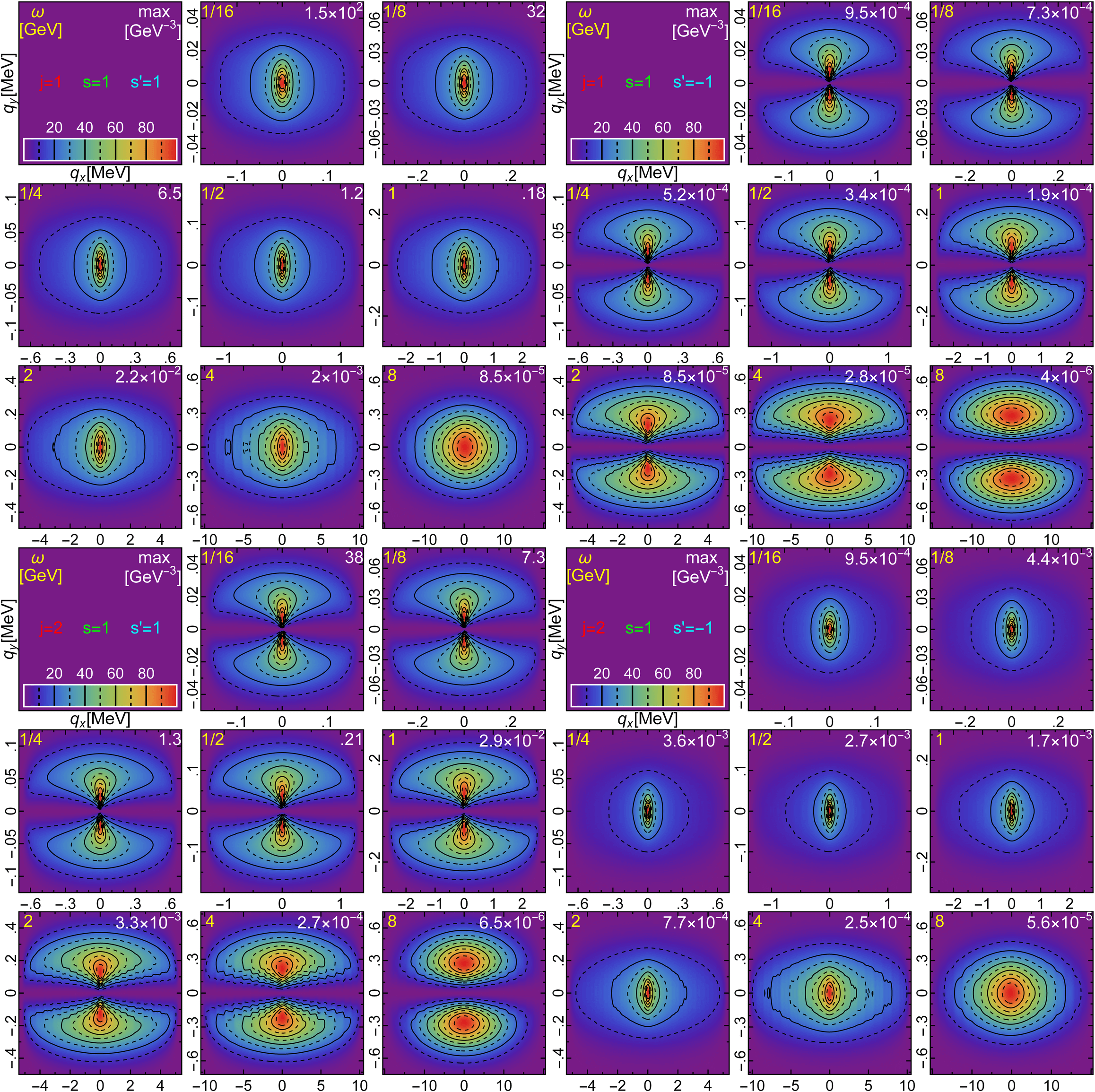}
\caption{Sections through the nonlinear Compton scattering probability distribution in GeV$^{-3}$, for $\tau=20\;\text{fs}$, $\varepsilon=10 \;\text{GeV}$, $\xi_0=50$, and incoming electron spin quantum number $s=1$. The color levels indicate percentages of the maximum reached by the distribution in that section. The approximate value of that maximum is shown at the upper right corner of each subplot.}
\label{Figure_7}
\end{center}
\end{figure}

\begin{figure}
\begin{center}
\includegraphics[width=1\columnwidth]{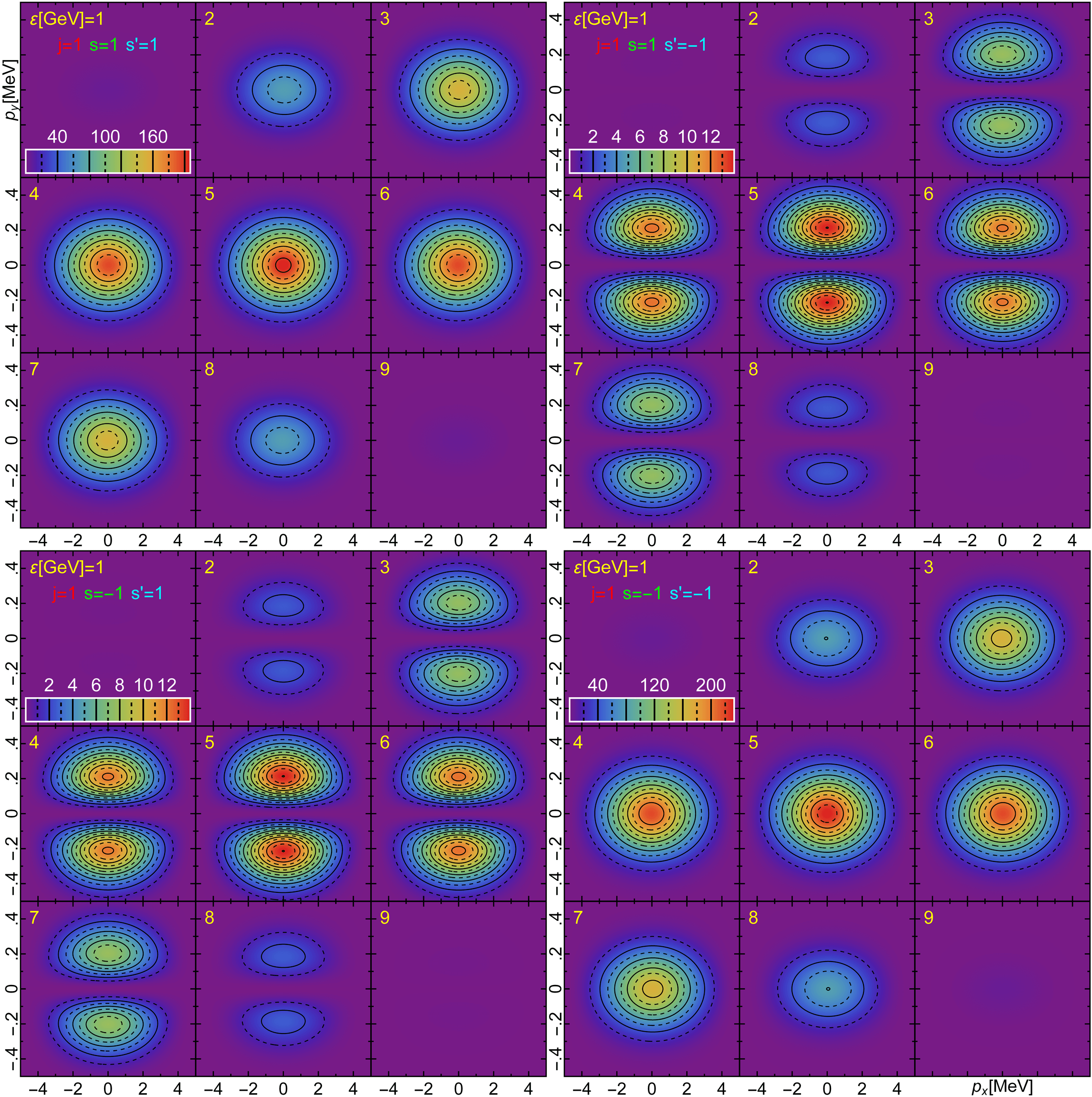}
\caption{Sections through the nonlinear Breit-Wheeler pair production probability distribution in GeV$^{-3}$, for $\tau=20\;\text{fs}$, $\omega=10\;\text{GeV}$, $\xi_0=10$, and incoming photon polarization state $j=1$. The horizontal axes correspond to $p_x$ [MeV] and the vertical ones correspond to $p_y$ [MeV].}
\label{Figure_8}
\end{center}
\end{figure}

\begin{figure}
\begin{center}
\includegraphics[width=1\columnwidth]{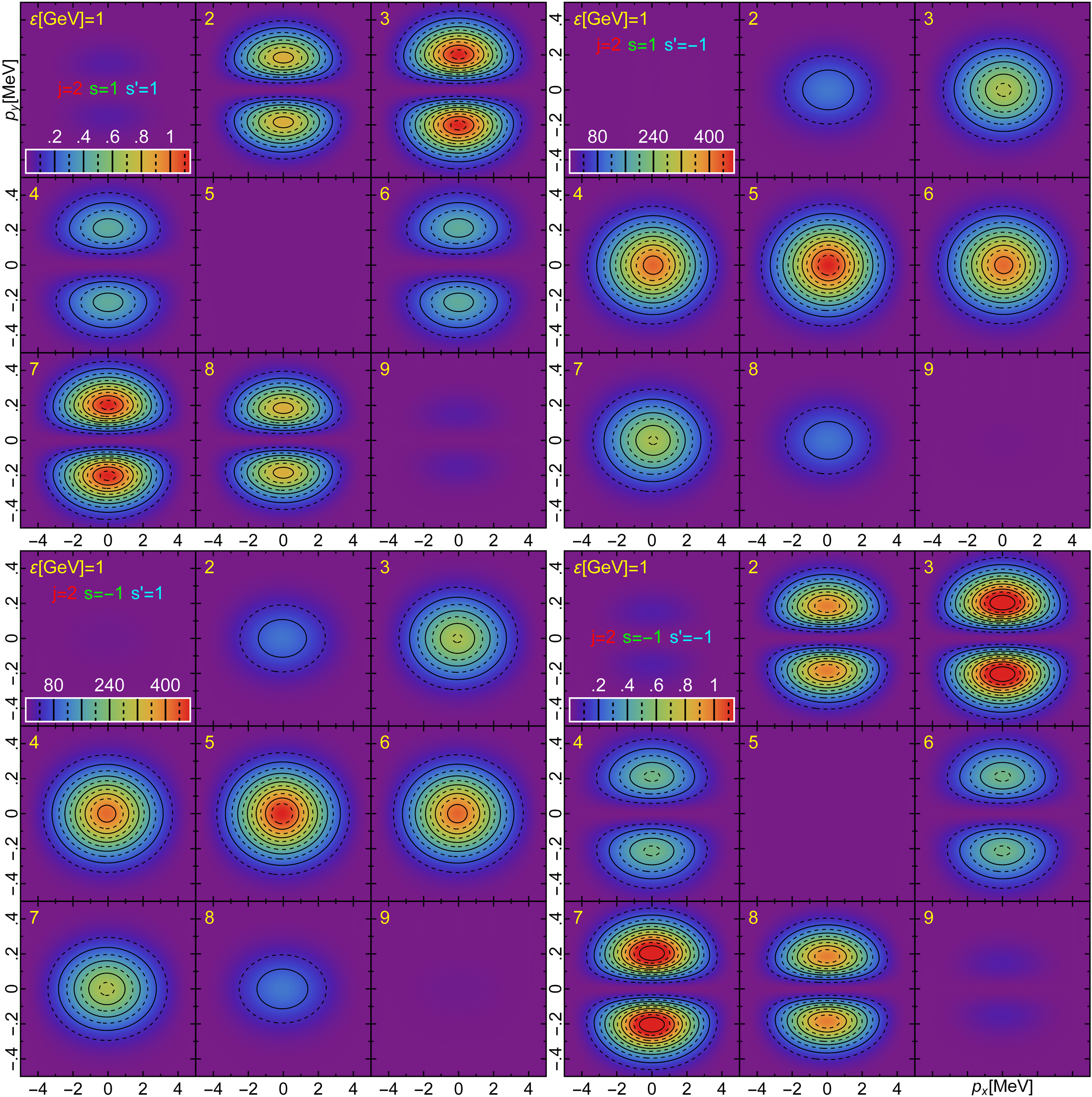}
\caption{Same as in Fig. \ref{Figure_8}, but for incoming photon polarization state $j=2$.}
\label{Figure_8bis}
\end{center}
\end{figure}

\begin{figure}
\begin{center}
\includegraphics[width=1\columnwidth]{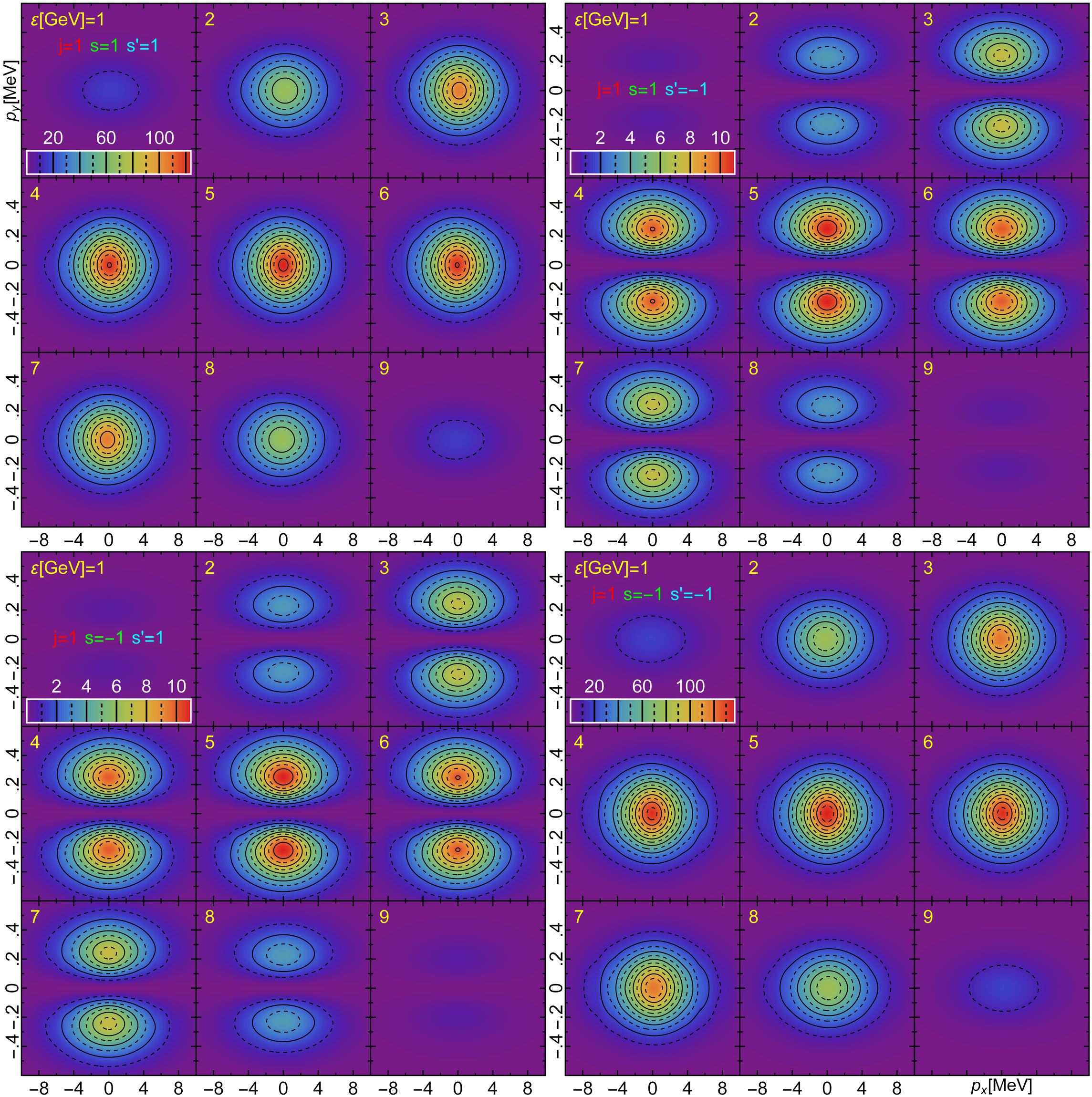}
\caption{Sections through the nonlinear Breit-Wheeler pair production probability distribution in GeV$^{-3}$, for $\tau=20\;\text{fs}$, $\omega=10\;\text{GeV}$, $\xi_0=50$, and incoming photon polarization state $j=1$. The horizontal axes correspond to $p_x$ [MeV] and the vertical ones correspond to $p_y$ [MeV].}
\label{Figure_9}
\end{center}
\end{figure}

\begin{figure}
\begin{center}
\includegraphics[width=1\columnwidth]{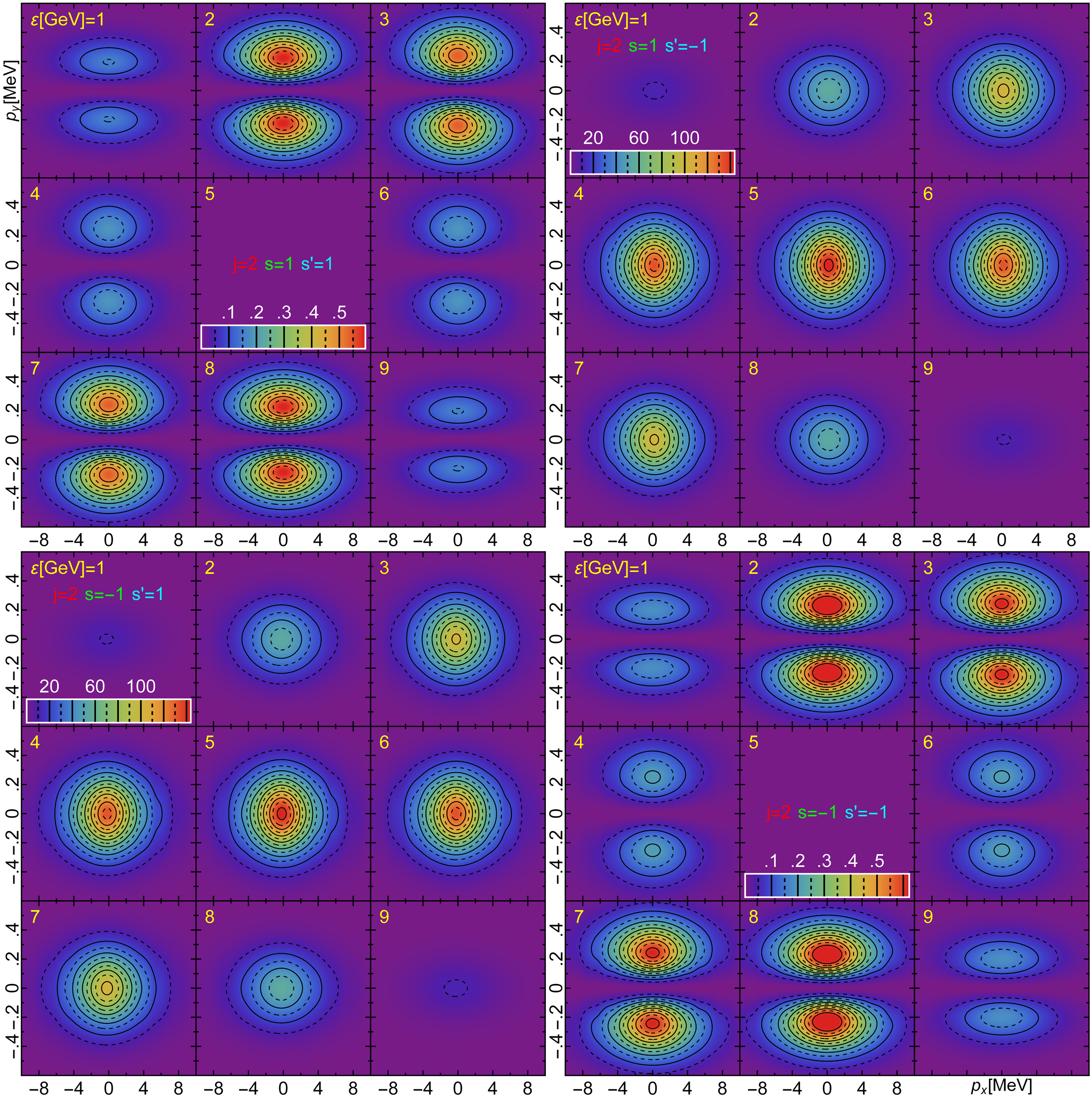}
\caption{Same as in Fig. \ref{Figure_9}, but for incoming photon polarization state $j=2$.}
\label{Figure_9bis}
\end{center}
\end{figure}

Now, we present sections of the fully differential distribution of the photon momentum for the nonlinear Compton scattering effect (Figs. \ref{Figure_6} and \ref{Figure_7}) and of the momentum of the positron produced in the nonlinear Breit-Wheeler pair production process (Figs. \ref{Figure_8} - \ref{Figure_9bis}). In all cases we fix the energy of the incoming particle to $10\ \text{GeV}$ (recall that the incoming particle is counterpropagating with respect to the plane wave) and we consider two laser intensities corresponding to $\xi_0=10$ and $50$, which are equivalent to $2.1\times 10^{20}\ \text{W}/\text{cm}^2$ and $5.4\times 10^{21}\ \text{W}/\text{cm}^2$, respectively. The sections are plotted as functions of $q_x=-q_1$ and $q_y=-q_2$ for nonlinear Compton scattering and of $p_x=-p_1$ and $p_y=-p_2$ for nonlinear Breit-Wheeler pair production, by keeping the light-cone energy of the emitted photon and of the produced positron fixed (which also means, with very good approximation, to keep fixed $q_3$ or $q_-$ and $p_3$ or $p_-$, respectively). Indeed, the plots show that the outgoing particles are most probably produced with transverse momenta much smaller than $|q_3|$ and $|p_3|$, respectively (see that the scale of the sections is in MeV units, instead of the GeV units used for the longitudinal momentum). As expected, the sections are asymmetric, spreading more along the direction of the electric field of the plane wave, i.e., the $x$ direction.

Both in the case of nonlinear Compton scattering and nonlinear Breit-Wheeler pair production, for some quantum number combinations the spectra vanish if the component of the momentum of the particle along the magnetic field of the wave, i.e., along the $y$ direction, vanishes. This can be explained analytically looking at Eqs. (\ref{T_prime_1}) and (\ref{T_prime_2}) for nonlinear Compton scattering and at Eqs. (\ref{G_NBW_1}) and (\ref{G_NBW_2}) for nonlinear Breit-Wheeler pair production. These equations show that for both processes the probabilities vanish in the cases $(j=1, s=-s')$ and $(j=2, s=s')$ for $q_2=0$ and $p_2=0$, respectively. Now, since the incoming particle is assumed to counterpropagate with respect to the plane wave for both processes, we have that $p_2=0$ in nonlinear Compton scattering and $q_2=0$ in nonlinear Breit-Wheeler pair production. Thus, if in addition $q_2=0$ in nonlinear Compton scattering and $p_2=0$ in nonlinear Breit-Wheeler pair production, the corresponding probability vanishes for the mentioned spin and polarization combinations. Furthermore, we see in Figs. \ref{Figure_6} - \ref{Figure_9bis} that the maximum of each section is found at the origin for the complementary combinations $(j=1, s=s')$ and $(j=2, s=- s')$. For the examples of nonlinear Breit-Wheeler pair production processes of Figs. \ref{Figure_8} - \ref{Figure_9bis} the aforementioned cases $(j=1, s=s')$ and $(j=2, s=-s')$ also have an altogether larger probability than the complementary ones $(j=1, s=-s')$ and $(j=2, s=s')$, as it can be recognized also noticing the different scales of the panels.

For nonlinear Compton scattering we only show results for the initial spin state defined by $s=1$ (see Figs. \ref{Figure_6} and \ref{Figure_7}) as the $s=-1$ case gives very similar results, provided one changes the sign of $s'$, too. The different scales in the transverse photon momenta for each individual plot show a general tendency of broadening of the probability distribution on the transverse plane for increasing energies of the emitted photon.

Finally, for nonlinear Breit-Wheeler pair production the probability density always vanishes for the energy of any final particle approaching the energy of the incoming photon. In addition, for ($j=2$, $s=s'$) it also vanishes when the final particles both have half of the total available energy (see Figs. \ref{Figure_8bis} and \ref{Figure_9bis}).

\section{Conclusions}

In conclusion, we presented analytical expressions and numerical evaluations of the probabilities for nonlinear Compton scattering and nonlinear Breit-Wheeler pair production including the particle states decay. The probabilities take into account that in a plane-wave background field the electron and photon states are not stable because electrons and positrons emit photons and because photons themselves decay into electron-positron pairs. Within our approach based on the locally-constant field approximation, the decay of the states leads to an exponential damping term in the expressions of the probabilities, which depends on the plane-wave light-cone time as well as on the light-cone energies and discrete quantum numbers of the participating particles.

In the analytical part, we first calculated the spin- and polarization-resolved traces and took the integrals over the transverse momenta and phase differences. The final probabilities depend on the spin and polarization quantum numbers, on the incoming particle four-momentum, as well as on the quantum nonlinearity parameter. Furthermore, we computed the asymptotic expressions for the probabilities differential in the light-cone momentum of one final particle for arbitrary spin and polarization quantum numbers in the limit of one of the outgoing particles gaining all or none of the light-cone energy of the incoming particle. All calculations were carried out using particular spin and polarization four-vectors and we proved that the results for the total probabilities are independent of the chosen spin and polarization bases.

In the numerical part we presented plots of the total and differential probabilities for nonlinear Compton scattering and nonlinear Breit-Wheeler pair production by considering two different pulse lengths. Due to the particle states decay, the total probabilities stay below unity in all cases. The damping becomes important for $\alpha \xi_0 \Phi_L \gtrsim 1$ at values of the quantum nonlinearity parameter of the order of unity, such that it increases more rapidly with $\xi_0$ for larger values of the pulse phase length $\Phi_L$. For nonlinear Compton scattering we saw that the probability behaves like a Poissonian distribution for low values of $\eta_0$ such that photon recoil is negligible, whereas important differences arise when $\eta_0$ increases. For the differential probability we found it has its maximum at vanishing perpendicular momentum in the case of polarization $j=1$ and same spins $s=s'$, and in the case of polarization $j=2$ and opposite spins $s=-s'$, otherwise it vanishes at $q_y=0$. The same behavior is observed for nonlinear Breit-Wheeler pair production for the same spin and polarization combination and for $p_y=0$, i.e., for the component of the positron momentum along the magnetic field of the wave. For nonlinear Breit-Wheeler pair production we have also seen that the differential probability vanishes for all spin and polarization combinations if the light-cone energy of the incoming photon all goes to either the electron or the positron. Also, for ($j=2$, $s=s'$) the differential probability vanishes if the electron and the positron have the same light-cone energy, i.e., half the light-cone energy of the incoming photon. Finally, in all the cases we investigated numerically, the kinematic conditions were such that 1) the transverse momenta of produced particles are much smaller than the corresponding longitudinal momentum and 2) the main spread of the transverse momenta is along the direction of the plane-wave electric field.

\
\acknowledgments{This publication was also supported and T. P. is funded by the Collaborative Research Centre 1225 funded by Deutsche Forschungsgemeinschaft (DFG, German Research Foundation) - Project-ID 273811115 - SFB 1225 
and for this work V. D.  was supported by a grant from the Romanian National Authority for Scientific Research, CNCS-UEFISCDI, the Project no. PN-III-P4-ID-PCCF-2016-0164.
Furthermore, this article comprises parts of the Ph.D. thesis work of T. P., submitted to Heidelberg University, Germany.
}

\begin{appendix}

\section{Computation of the trace}

We present here the computation of the traces for nonlinear Compton scattering and nonlinear Breit-Wheeler pair production given in Eq. (\ref{Trace}) and Eq. (\ref{Trace_NBW}), respectively. For this we first simplify the expressions by introducing the quantity
\begin{equation}\begin{split}
Q_{p,s}(\phi_+,\phi_-)=&\left[1+\frac{\hat{n}[\hat{\mathcal{\mathcal{A}}}(\phi_+)+\hat{\mathcal{A}}'(\phi_+)\phi_-/2]}{2p_-}\right] (\hat{p}+m) (1+s\gamma^5 \hat{\zeta})\\
&\times \left[1-\frac{\hat{n}[\hat{\mathcal{\mathcal{A}}}(\phi_+)-\hat{\mathcal{A}}'(\phi_+)\phi_-/2]}{2p_-}\right].
\end{split}\end{equation}
With this notation the trace for nonlinear Compton scattering is given by
\begin{equation}\label{T_NCS_gamma}
T_{j,s,s'}= \frac{1}{4} \text{tr} \left\{ \hat{\Lambda}_j(q) Q_{p,s}(\phi_+,\phi_-) \hat{\Lambda}_j(q) Q_{p',s'}(\phi_+,-\phi_-) \right\}
\end{equation}
and for nonlinear Breit-Wheeler pair production the trace reads
\begin{equation}\label{T_NBW_gamma}
G_{j,s,s'}= -\frac{1}{4} \text{tr} \left\{ \hat{\Lambda}_j(q) Q_{-p,s}(\phi_+,\phi_-) \hat{\Lambda}_j(q) Q_{p',s'}(\phi_+,-\phi_-) \right\}.
\end{equation}
Now the function $Q_{p,s}(\phi_+,\phi_-)$ can be decomposed into a linear combination of the matrices $1_{4\times 4}$, $\gamma^5$, $\gamma^{\mu}$, $i\gamma^{\mu}\gamma^5$, $\sigma^{\mu\nu}=(i/2) (\gamma^{\mu} \gamma^{\nu} -\gamma^{\nu} \gamma^{\mu})$ such that 
\begin{equation} \label{SpinBasis}
Q_{p,s}(\phi_+,\phi_-)=c_1 1_{4\times 4} +c_5 \gamma^5 +c_{\mu} \gamma^{\mu} +c_{5\mu} i\gamma^{\mu}\gamma^5 +c_{\mu\nu} \sigma^{\mu\nu},
\end{equation}
where the coefficients are obtained by solving the following traces
\begin{align}
\label{cal_c1}
c_1=& \frac{1}{4} \text{Tr}\left[ 1_{4\times 4} Q_{p,s}(\phi_+,\phi_-)\right],\\
c_5=& \frac{1}{4} \text{Tr}\left[ \gamma^5 Q_{p,s}(\phi_+,\phi_-)\right],\\
c_{\mu}=& \frac{1}{4} \text{Tr}\left[ \gamma_{\mu} Q_{p,s}(\phi_+,\phi_-)\right],\\
c_{5\mu}=& \frac{1}{4} \text{Tr}\left[ i\gamma_{\mu}\gamma^5 Q_{p,s}(\phi_+,\phi_-)\right],\\
\label{cal_cmunu}
c_{\mu\nu}=& \frac{1}{8} \text{Tr}\left[ \sigma_{\mu\nu} Q_{p,s}(\phi_+,\phi_-)\right].
\end{align}
These traces can be calculated and we present them here for a linear polarized plane wave background field
\begin{equation}\label{c_1}
c_1= m -i\frac{s}{4p_-} \epsilon^{\alpha \beta \gamma \delta} \zeta_{\alpha}  \mathcal{F}_{\beta\gamma} \psi'(\phi_+) \phi_- p_{\delta},
\end{equation}
\begin{equation}\label{c_5}
c_5= 0,
\end{equation}
\begin{equation}\label{c_mu}\begin{split}
c_{\mu}=& p_{\mu} -\mathcal{A}_{\mu}(\phi_+) -\frac{1}{p_-} n_{\mu} \bm{p}_{\perp} \cdot \bm{\mathcal{A}}_{\perp}(\phi_+)\\
 &+\frac{1}{2p_-} n_{\mu} \left[\bm{\mathcal{A}}_{\perp}^2(\phi_+) -\bm{\mathcal{A}}'^2_{\perp}(\phi_+) \frac{\phi_-^2}{4}\right]\\
 &-i \frac{ms}{4p_-} \eta_{\mu\nu} \epsilon^{\alpha \beta \delta \nu} \zeta_{\alpha} \mathcal{F}_{\beta\delta} \psi'(\phi_+) \phi_-,
\end{split}\end{equation}
\begin{equation}\label{c_5mu}\begin{split}
c_{5\mu}=& ims \zeta_{\mu} +\frac{1}{4p_-} \eta_{\mu\nu} \epsilon^{\alpha \beta \delta \nu} p_{\alpha} \mathcal{F}_{\beta\delta} \psi'(\phi_+) \phi_- ,
\end{split}\end{equation}
and
\begin{equation}\label{c_munu}\begin{split}
c_{\mu\nu}=& \frac{1}{2} \bigg\{ -i\frac{m}{2p_-} \mathcal{F}_{\mu\nu} \psi'(\phi_+) \phi_- 
-s \epsilon_{\mu\nu\rho\sigma} p^{\rho} \zeta^{\sigma} \\
&+\frac{s}{2p_-} (\eta_{\nu\rho} \epsilon_{\mu\sigma\tau\delta} -\eta_{\mu\rho} \epsilon_{\nu\sigma\tau\delta}) p^{\rho} \zeta^{\sigma} \mathcal{F}^{\tau\delta} \psi(\phi_+)\\
&+\frac{s}{2p_-} \epsilon_{\mu\nu\tau\rho} n^{\tau} \zeta^{\rho} \left[\mathcal{A}_{\delta}(\phi_+) \mathcal{A}^{\delta}(\phi_+) -\mathcal{A}'_{\delta}(\phi_+) \mathcal{A}'^{\delta}(\phi_+) \frac{\phi_-^2}{4}\right] \bigg\}.
\end{split}\end{equation}

The total trace of nonlinear Compton scattering, given in Eq. (\ref{T_NCS_gamma}), contains also the function $Q_{p',s'}(\phi_+,-\phi_-)$ for which we introduce the notation
\begin{equation}
Q_{p',s'}(\phi_+,-\phi_-)=c_1' 1_{4\times 4} +c_5' \gamma^5 +c_{\tau}' \gamma^{\tau} +c_{5\tau}' i\gamma^{\tau}\gamma^5 +c_{\tau\lambda}' \sigma^{\tau\lambda},
\end{equation}
where the primed coefficients are obtained analogously to Eqs. (\ref{cal_c1})-(\ref{cal_cmunu}).
With this the trace for nonlinear Compton scattering can be reduced to the form
\begin{equation}\begin{split}\label{T_total_NCS}
T_{j,s,s'}=&-c_1 c_1' +c_5 c_5' +(c_{\mu} c_{\tau}' -c_{5\mu} c_{5\tau}') (2 \Lambda_j^{\mu}(q) \Lambda_j^{\tau}(q) +\eta^{\mu\tau})\\ 
&-2 c_{\mu\nu} c_{\tau\lambda}' \eta^{\nu\lambda} (4\Lambda_j^{\mu}(q) \Lambda_j^{\tau}(q) +\eta^{\mu\tau}).
\end{split}\end{equation}
The trace turns out to only depend on contractions of the primed and the corresponding not primed coefficients. These contractions can be calculated and they are given by
\begin{equation}\label{cc_1}\begin{split}
c_1c'_1=& m^2 -i\frac{m}{4} \epsilon^{\alpha \beta \gamma \delta} \mathcal{F}_{\beta\gamma} \psi'(\phi_+) \phi_- \left[ \frac{s}{p_-} \zeta_{\alpha} p_{\delta} - \frac{s'}{p'_-} \zeta'_{\alpha} p'_{\delta} \right]\\
& +\frac{1}{16} \frac{ss'}{p_- p'_-} \epsilon^{\alpha \beta \gamma \delta} \mathcal{F}_{\beta \gamma} \zeta_{\alpha} p_{\delta} \epsilon^{\alpha' \beta' \gamma' \delta'} \mathcal{F}_{\beta' \gamma'} \zeta'_{\alpha'} p'_{\delta'} \psi'^2 (\phi_+) \phi_-^2,
\end{split}\end{equation}
\begin{equation}
c_5 c'_5= 0,
\end{equation}
\begin{equation}\begin{split}
c_{\mu} c'_{\tau} &\left(2 \Lambda_j^{\mu}(q) \Lambda_j^{\tau}(q) +\eta^{\mu \tau}\right)=
(p p') +(\bm{p}_{\perp} +\bm{p}'_{\perp})\cdot \bm{\mathcal{A}}_{\perp}(\phi_+)\\
&- \left[\frac{p'_-}{p_-} \bm{p}_{\perp}+\frac{p_-}{p'_-}\bm{p}'_{\perp} \right] \cdot \bm{\mathcal{A}}_{\perp}(\phi_+) 
+\frac{1}{2} \left(\frac{p'_-}{p_-}+\frac{p_-}{p'_-}\right) \left[\bm{\mathcal{A}}_{\perp}^2(\phi_+) -\bm{\mathcal{A}}'^2_{\perp}(\phi_+) \frac{\phi_-^2}{4} \right]\\
&+\mathcal{A}_{\mu}(\phi_+) \mathcal{A}^{\mu}(\phi_+)
-i\frac{m}{4} \epsilon^{\alpha \beta \gamma \delta} \mathcal{F}_{\beta\gamma} \psi'(\phi_+) \phi_- \left[ \frac{s}{p_-} \zeta_{\alpha} p'_{\delta} - \frac{s'}{p'_-} \zeta'_{\alpha} p_{\delta} \right]\\
&+2 \left(p \Lambda_j(q)\right) \left(p' \Lambda_j(q)\right) 
-2[p_{\mu} + p'_{\mu}] \Lambda_j^{\mu}(q) \left( \Lambda_j(q) \mathcal{A}(\phi_+) \right)\\
&+2\left(\Lambda_j(q) \mathcal{A}(\phi_+)\right)^2,
\end{split}\end{equation}
\begin{equation}\begin{split}
c_{5\mu} c'_{5\tau} &\left(2 \Lambda_j^{\mu}(q) \Lambda_j^{\tau}(q) +\eta^{\mu \tau}\right)=
-m^2 s s' \left[2 \left( \zeta \Lambda_j(q) \right) \left( \zeta' \Lambda_j(q) \right) +(\zeta \zeta') \right]\\
&+i \frac{m}{4} \epsilon^{\alpha \beta \gamma \delta} \mathcal{F}_{\beta\gamma} \psi'(\phi_+) \phi_- \left[ \frac{s'}{p_-} p_{\alpha} \zeta'_{\delta} - \frac{s}{p'_-} p'_{\alpha} \zeta_{\delta} \right]\\
&+i\frac{m}{2} \epsilon^{\alpha \beta \gamma \delta} \mathcal{F}_{\beta\gamma} \psi'(\phi_+) \phi_- \left[\frac{s'}{p_-} p_{\alpha} \left(\zeta'\Lambda_j(q)\right) -\frac{s}{p'_-} p'_{\alpha} \left(\zeta \Lambda_j(q)\right) \right] \Lambda_{j,\delta}(q)\\
&- \frac{1}{16p_- p'_-} \epsilon^{\alpha \beta \gamma \delta} p_{\alpha} \mathcal{F}_{\beta \gamma} \epsilon^{\alpha' \beta' \gamma' \delta'} p'_{\alpha'} \mathcal{F}_{\beta' \gamma'} \left[\eta_{\delta \delta'} +2\Lambda_{j,\delta}(q) \Lambda_{j,\delta'}(q)\right] \psi'^2 (\phi_+) \phi_-^2,
\end{split}\end{equation}
and
\begin{equation}\label{cc_munu}\begin{split}
2 c_{\mu\nu} &c'_{\tau\lambda} \eta^{\nu\lambda} \left( 4 \Lambda_j^{\mu}(q) \Lambda_j^{\tau}(q) +\eta^{\mu\tau} \right) = - s s' \left[(p \zeta') (p' \zeta)-(p p') (\zeta \zeta')\right]\\
&-2 ss' \Big[(p\zeta')(p'\Lambda_j(q))(\zeta \Lambda_j(q))+(\zeta p')(\zeta'\Lambda_j(q))(p\Lambda_j(q))\\ &\qquad \qquad -(pp')(\zeta \Lambda_j(q))(\zeta'\Lambda_j(q))-(\zeta \zeta')(p\Lambda_j(q))(p'\Lambda_j(q))\Big] \\
&+i \frac{m}{4} \psi'(\phi_+) \phi_- \epsilon^{\mu\nu\rho\sigma} \left[ \mathcal{F}_{\mu\nu} +4 \Lambda_{j,\mu}(q) \left( \Lambda_{j}^{\tau}(q) \mathcal{F}_{\tau\nu} \right) \right] \left[ \frac{s'}{p_-} p'_{\rho} \zeta'_{\sigma} - \frac{s}{p'_-} p_{\rho} \zeta_{\sigma} \right]\\
&-\frac{ss'}{2} \left[\mathcal{A}_{\delta}(\phi_+) \mathcal{A}^{\delta}(\phi_+) -\mathcal{A}'_{\delta}(\phi_+) \mathcal{A}'^{\delta}(\phi_+) \frac{\phi_-^2}{4}\right] \\
&\qquad\times \left[ (\zeta \zeta')+ 2 (\zeta \Lambda_j(q)) (\zeta'\Lambda_j(q))\right] \left(\frac{p_-}{p'_-}+\frac{p'_-}{p_-} \right) \\
&+ss' \psi(\phi_+) (\zeta \zeta') \left[\frac{p_- -p'_-}{p_- p'_-} (p' \mathcal{F} p) -2 \frac{(p'\Lambda_j(q))}{p'_-} (p\mathcal{F}\Lambda_j(q)) -2 \frac{(p\Lambda_j(q))}{p_-} (p'\mathcal{F}\Lambda_j(q))\right] \\
&-\frac{ss'}{4p_- p'_-} \psi^2(\phi_+) (\epsilon^{\rho'\sigma\gamma\delta} p'_{\rho'} \zeta_{\sigma} \mathcal{F}_{\gamma\delta}) (\epsilon^{\rho\sigma'\gamma'\delta'} p_{\rho} \zeta'_{\sigma'} \mathcal{F}_{\gamma'\delta'}).
\end{split}\end{equation}
Inserting Eqs. (\ref{cc_1})-(\ref{cc_munu}) into Eq. (\ref{T_total_NCS}) and simplifying the expression for the two polarization states $j=1$ and $j=2$ leads finally to Eqs. (\ref{T_1}) and (\ref{T_2}), respectively.

The trace for nonlinear Breit-Wheeler pair production given in Eq. (\ref{T_NBW_gamma}) can be derived directly from the result of nonlinear Compton scattering in Eqs. (\ref{T_total_NCS})-(\ref{cc_munu}) by multiplying Eq. (\ref{T_total_NCS}) by an overall minus sign and by changing the sign of the four-momentum $p^{\mu}$ in the Eqs. (\ref{cc_1})-(\ref{cc_munu}). With that one finally obtains Eqs. (\ref{G_NBW_1}) and (\ref{G_NBW_2}) for the two polarization states $j=1$ and $j=2$, respectively.

\end{appendix}

\end{document}